\documentclass{emulateapj}

\journalinfo{To Appear in ApJ}
\slugcomment{Preprint version \today}

\shorttitle{The effective temperature scale of FGK stars. II.}

\shortauthors{Ram\'{\i}rez \& Mel\'endez}

\newcommand{\teff}{T_\mathrm{eff}}
\newcommand{\feh}{\mathrm{[Fe/H]}}
\newcommand{\logg}{\log g}

\begin{document}

\title{The effective temperature scale of FGK stars. II. $\teff:\mathrm{color}:\feh$ calibrations}

\author{Iv\'an Ram\'{\i}rez\altaffilmark{1}}
\affil{Department of Astronomy, University of Texas at Austin, RLM 15.306, TX 78712-1083} \and
\author{Jorge Mel\'endez\altaffilmark{1}} \affil{Department of Astronomy, California Institute of Technology, MC 105--24, Pasadena, CA 91125}

\altaffiltext{1}{Affiliated with the Seminario Permanente de Astronom\'ia y Ciencias Espaciales of the Universidad Nacional Mayor de San Marcos, Peru}

\begin{abstract}
We present up-to-date metallicity-dependent temperature vs. color calibrations for main sequence and giant stars based on temperatures derived with the infrared flux method (IRFM). Seventeen colors in the following photometric systems: $UBV$, $uvby$, Vilnius, Geneva, $RI$(Cousins), DDO, Hipparcos-Tycho, and 2MASS, have been calibrated. The spectral types covered range from F0 to K5 (7000~K$\gtrsim\teff\gtrsim$4000~K) with some relations extending below 4000~K or up to 8000~K. Most of the calibrations are valid in the metallicity range $-3.5\gtrsim\feh\gtrsim0.4$, although some of them extend to as low as $\feh\sim-4.0$. All fits to the data have been performed with more than 100 stars; standard deviations range from 30~K to 120~K. Fits were carefully performed and corrected to eliminate the small systematic errors introduced by the calibration formulae. Tables of colors as a function of $\teff$ and $\feh$ are provided. This work is largely based on the study by A. Alonso and collaborators and thus our relations do not significantly differ from theirs except for the very metal-poor hot stars. From the calibrations, the temperatures of 44 dwarf and giant stars with direct temperatures available are obtained. The comparison with direct temperatures confirms our finding in Part~I that the zero point of the IRFM temperature scale is in agreement, to the 10~K level, with the absolute temperature scale (that based on stellar angular diameters) within the ranges of atmospheric parameters covered by those 44 stars. The colors of the Sun are derived from the present IRFM $\teff$ scale and they compare well with those of five solar analogs. It is shown that if the IRFM $\teff$ scale accurately reproduces the temperatures of very metal-poor stars, systematic errors of the order of 200~K, introduced by the assumption of $(V-K)$ being completely metallicity-independent when studying very metal-poor dwarf stars, are no longer acceptable. Comparisons with other $\teff$ scales, both empirical and theoretical, are also shown to be in reasonable agreement with our results, although it seems that both Kurucz and MARCS synthetic colors fail to predict the detailed metallicity dependence, given that for $\feh=-2.0$, differences as high as $\sim\pm200$~K are found. 
\end{abstract}


\section{Introduction}

Theoretical calculations in astrophysics predict relations between physical quantities such as effective temperature ($\teff$), luminosity ($L$), and stellar radius. These quantities are, in general, not directly measurable, and so their correspondence with observational quantities such as colors and magnitudes play a crucial role in the interpretation of the results. The importance and usefulness of such correlations are discussed in the following paragraphs.

Stellar chemical compositions are derived from the comparison of synthetic and observed spectra, either by line-profile fitting (e.g. Hill et~al. 2002, Allende Prieto et~al. 2004, Sneden et~al. 2003) or equivalent width matches (e.g. Reddy et~al. 2003, Takeda et~al. 2002). Both methods require both $\teff$ and $\logg$ (the surface gravity) as input parameters and so their uncertainties are reflected in the abundances derived. The accurate determination of effective temperatures is thus a critical step in any abundance analysis. 

The theory of stellar evolution deals with the evolution in time of the fundamental stellar physical parameters, which is very well illustrated in the $\teff$ vs. $L$ plane (the theoretical HR diagram). Theoretical isochrones and evolutionary tracks are the final products of these calculations (e.g. Girardi et~al. 2002, Yi et~al. 2003). The transformation of the effective temperature axis into a color axis, along with a transformation of the luminosity axis into an absolute magnitude axis (i.e. the transformation of the $\teff$ vs. $L$ plane into a color-magnitude diagram), allow observations to be compared with theoretical results, leading to a better understanding of the systems studied, or to the test of the models themselves. There is a continuous feedback between theory and observation through these kinds of transformations.

A problem of particular interest, whose resolution may be partly in the adopted $\teff$ scale is that of the primordial lithium abundance, $A$(Li), which is derived from the observation of metal-poor dwarfs. Ryan et~al. (1999) determined lithium abundances in very metal-poor stars ($-3.6<\feh<-2.3$) employing the temperature calibration of Magain (1987), whose metallicity dependence was derived with only one star (HD 140283, $\feh=-2.5$) in the metallicity range used by Ryan et~al., and the few other metal-poor stars included in the calibration have $\feh>-2.15$. Ryan et~al. claim that the lithium abundance in halo stars depends on metallicity, and extrapolating to zero metals they derive a primordial abundance of $A$(Li$)=2.0$ dex. On the other hand, using the standard theory of Big Bang nucleosynthesis and the baryon-to-photon ratio determined from WMAP (Wilkinson Microwave Anisotropy Probe), a primordial lithium abundance of 2.6 dex is derived (Romano et~al. 2003, Coc et~al. 2004), much higher than the lithium abundance obtained by Ryan et~al. (1999).

Bonifacio \& Molaro (1997) found that when the $\teff$ scale from the infrared flux method (IRFM) is employed, the lithium abundance in very metal-poor stars does not depend on metallicity, and a primordial $A$(Li) = 2.24 dex is obtained, significantly higher than the primordial abundance proposed by Ryan et~al., but still lower than the abundance suggested by the WMAP data.
Since the temperature is the key parameter to obtain lithium abundances, we have included very metal-poor F and G dwarfs in our calibrations, in order to diminish the largest source of uncertainty in the Li controversy. In fact, a reanalysis of the Li spectroscopic data using the present temperature scale leads to a Li plateau with $A$(Li$)=2.37$ (Mel\'endez \& Ram\'{\i}rez 2004), a value that is closer to that suggested by WMAP. A discrepancy, although much smaller than that reported in previous works, still persists.

The present work aims to a better definition of the temperature vs. color relations, taking into account the effect of the different chemical compositions (as measured by the metallicity $\feh$) observed in the atmospheres of F, G, and K stars. In the first part of this work (hereafter Part~I), we derived the temperatures of about a thousand stars with the IRFM. Combining these results with photometric measurements of the sample stars we now proceed to calibrate the $\teff:\mathrm{color}:\feh$ relations in the following photometric systems (the corresponding colors are given in parenthesis): $UBV$ ($B-V$), $uvby$ ($b-y$), Vilnius ($Y-V$, $V-S$), Geneva ($B_2-V_1$, $B_2-G$, $t$), Johnson-Cousins ($V-R_C$, $V-I_C$, $R_C-I_C$), DDO ($42-45$, $42-48$), Johnson-2MASS ($V-J_2$, $V-H_2$, $V-K_2$), Tycho ($B_T-V_T$), and Tycho-2MASS $(V_T-K_2)$.

This paper revisits the widely used results of Alonso et~al. (1996, 1999; hereafter AAM), as well as our earlier extensions (Mel\'endez \& Ram\'{\i}rez 2003, hereafter MR03; Ram\'{\i}rez \& Mel\'endez 2004a, hereafter RM04a). In \S\ref{sect:sample} the characteristics of the sample adopted (better described in Part~I) are given along with the sources of the photometry. The nature of the calibration formulae and the fits to the data are described in \S\ref{sect:fits}. The empirical temperature scale is tabulated and tested in \S\ref{sect:scale}. The comparison with other $\teff$ scales is given in \S\ref{sect:comparison} and the conclusions are summarized in \S\ref{sect:conclusions}.

\section{The sample, temperatures and photometry adopted} \label{sect:sample}

Approximately 80\% of the sample we used to calibrate the color-$\teff$ relations come from AAM work. The stellar metallicities, however, have been assigned according to the `2003 updated' Cayrel de Strobel et~al. (2001) catalog or the photometric calibration we derived in Part~I, which is also based on this catalog. The remaining stars are from a sample of planet-hosting stars (Ram\'{\i}rez \& Mel\'endez 2004b), and a selected sample of metal-poor and metal-rich dwarf and giant stars. The latter allow a better coverage of the regions below $\feh\sim-3.0$ and above $\feh\sim0.1$, as well as the dwarf metal-poor cool end ($\teff\sim4500$ K), with reliable input data. The stars for which kinematical metallicities were adopted in Part~I were excluded from the calibrations. See Sect.~3.2 in Part~I for details and references.

Effective temperatures were derived in Part I from an implementation of the IRFM (see e.g. Blackwell et~al. 1980). There we showed that the IRFM temperatures, which have a mean uncertainty of 1.3\%, are very well scaled to the direct ones (those derived from angular diameter and bolometric flux measurements) for $\feh>-0.6$, and 4000~K$<\teff<6500$~K for dwarfs or 3800~K$<\teff<5000$~K for giants. For the rest of the atmospheric parameters space, we still rely on the capability of the Kurucz models (those adopted in Part~I for the IRFM implementation) to reproduce the low blanketing effects in the infrared ($>1\mu$m).

The temperatures we derived are not strictly consistent with the whole set of metallicities adopted since not all of the $\feh$ values given in the literature were derived using IRFM temperature scales, i.e. a redetermination of the iron abundances with our temperature scale would be needed in order to have a consistent set of $\teff$ and $\feh$. However, errors of 100 K in $\teff$ result in errors of about 0.05-0.10 dex in $\feh$ for both dwarfs (e.g. Reddy et~al. 2003, Gratton et~al. 2003) and giants (Shetrone 1996, Mel\'endez et~al. 2003, Francois et~al. 2003), which, in turn, may affect the $\teff$ by only about 10~K in a second iteration with the IRFM. Therefore, these small inconsistencies do not significantly have an impact on our IRFM calibrations. Furthermore, the adoption of the mean of several metallicity determinations as well as the use of hundreds of stars to define the $\teff:\mathrm{color}:\feh$ relations minimize the effect.

The IRFM implementation from Part~I uses essentially only the $V$ magnitude and the infrared photometry. The bolometric fluxes were obtained from calibrations that use only $K$ and $(V-K)$ and have an internal accuracy of 1\% whereas the systematic errors on the calibration used, if present, will not affect the temperatures by more than about a conservative estimate of 50~K (Sect.~3.4 in Part~I, see also Ram\'{\i}rez \& Mel\'endez 2004b). Photometric errors are easily propagated and are reflected in the IRFM temperatures. Despite this, whenever reliable photometry is adopted, the IRFM temperatures are accurate to within $\sim$75~K. Since we do not expect this to be the case for the whole sample, we strongly recommend the use of several of the color-temperature calibrations derived here, and give the mean value a larger weight than the temperature from the IRFM, where available. This reduces not only the errors in $\teff$ introduced by the photometry in the color-$\teff$ calibrations, but also the error due to the IRFM $\teff$.

$UBV$, $uvby$, Vilnius, Geneva, $RI$(Cousins), and DDO photometry has been taken from several catalogs included in the General Catalogue of Photometric Data (Mermilliod et~al. 1997, hereafter GCPD). Mean values of $(B-V)$ and $(b-y)$, as given in the GCPD, have been adopted. Due to the low number of giants with $RI$(Cousins) photometry available in the GCPD, we took Washington or Kron-Eggen photometry and put them into the Cousins system by means of the transformation equations of Bessell (1979, 2001). This is the only place where color-color transformations have been used, and so the calibrations for giants in the Cousins system must be taken with care. Note, however, that the filters involved are not very different from those of the Cousins system, especially for the Washington system. Photometry from the Hipparcos-Tycho mission (ESA 1997) was also used, as well as the infrared photometry from the final release of the Two Micron All Sky Survey (2MASS, Cutri et~al. 2003). We discuss each system in turn.

$UBV$ (Johnson \& Morgan 1953) and $uvby$ (Str\"omgren 1966) were selected because they are widely used. In order to use existing photometry of thousands of stars in other systems we also calibrated the visual and infrared colors in the Vilnius (Kararas et~al. 1966, see also Strai\v{z}ys \& Sviderskiene 1972) and Cousins (1976) systems, respectively. 

Geneva photometry (Golay 1966) is considered one of the major systems available. It is both very homogeneous and the number of stars observed in both hemispheres is large. We employed the $t\equiv(B_2-G)-0.39(B_1-B_2)$ parameter (Strai\v{z}ys 1995), whose metallicity sensitivity is not as strong as in other Geneva colors, in addition to the $(B_2-V_1)$ and $(B_2-G)$ colors.

Although the colors from the DDO system (McClure \& van~den~Bergh 1968) are very metallicity dependent and the number of stars observed is not particularly large, the extremely careful observations performed with this system allow the metallicity effects to be easily distinguished from photometric uncertainties, and so whenever reliable abundances (and DDO photometry) are available, accurate temperatures may be derived. Both $C(42-45)$ and $C(45-48)$ are satisfactory temperature indicators (besides the strong metallicity dependence), but the dispersion in the $\teff$ vs. $C(45-48)$ plane is too large, and so we have preferred to calibrate the $C(42-48)\equiv C(42-45)+C(45-48)$ color.

A very large number of stars has been observed with the all sky surveys of Hipparcos-Tycho and 2MASS. The first of these contains mainly bright stars ($V_T<12$) while the very bright star ($V<5$) photometry of the second one has lower quality and was not considered in general. Note, however, that the 2MASS photometry is still accurate for stars as faint as $V\sim14$, and those have been included in the present work. 2MASS photometry was adopted whenever the error in the $K$ magnitude was less than $0.025$.

Reddening corrections for $(B-V)$ were already given in Part I. The reddening ratios used to correct the remaining colors, $k=E(\mathrm{color})/E(B-V)$, are given in Table~\ref{t:red}. They were mainly obtained from Schlegel et~al. (1998) table of `relative extinction for selected bandpasses,' adopting the appropriate effective wavelengths of the filters, as given in the Asiago Database of Photometric Systems (Fiorucci \& Munari 2003, for $r=1.00$ and $E(B-V)=0$). Reddening ratios $k$(color) obtained in this way are in very good agreement with the values given or calculated from related results in the literature (Table~\ref{t:red}). 
For the Johnson-Cousins colors, however, interpolation from the Schlegel et~al. (1998) tables is not a safe procedure given that the effective wavelengths of the Cousins filters strongly depend on spectral type (Bessell 1986). Using the effective wavelengths given by Fiorucci \& Munari (2003) for $R_C$ and $I_C$ we obtain $k(V-R_C)=0.51$, $k(V-I_C)=1.12$, and $k(R_C-I_C)=0.62$ while the values often quoted in the literature are around 0.6, 1.3, and 0.7, respectively (Dean et~al. 1978, Taylor 1986, Bessell et~al. 1998). The latter values are preferred.

\begin{deluxetable*}{ccrcc}
\tablecaption{Adopted Extinction Ratios and Comparison with the Literature}
\tablehead{\colhead{Color} & System & \colhead{$k$\tablenotemark{1}} & \colhead{$k$(Literature)} & \colhead{Reference}}
\startdata
$(b-y)$ & Str\"omgren & 0.74 & 0.74 & Crawford (1975) \\
$(Y-V)$ & Vilnius & 0.72 & 0.74 & Kuriliene \& Sudzius (1974) \\
$(V-S)$ & Vilnius & 0.62 & 0.62 & Sudzius et al. (1996) \\
$(B_2-G)$ & Geneva & 1.14 & 1.14 & Bersier (1996) \\
$(B_2-V_1)$ & Geneva & 0.86 & 0.85 & Bersier (1996) \\
$t$ & Geneva & 0.98 & 0.99 & Bersier (1996) \\
$(V-R_C)$ & Johnson-Cousins & 0.60  & 0.60 & Taylor (1986) \\
$(V-I_C)$ & Johnson-Cousins & 1.30 & 1.25, 1.34 & Dean et al. (1978), Taylor (1986) \\
$(R_C-I_C)$ & Cousins & 0.70 & 0.70 & Taylor (1986) \\
$C(42-45)$ & DDO & 0.23 & 0.23 & Dawson (1978) \\
$C(42-48)$ & DDO & 0.58 & 0.59 & Dawson (1978) \\
$(V-J_2)$ & Johnson-2MASS & 2.16 & 2.25 & McCall (2004) \\
$(V-H_2)$ & Johnson-2MASS & 2.51 & 2.55 & McCall (2004) \\
$(V-K_2)$ & Johnson-2MASS & 2.70 & 2.72 & McCall (2004) \\
$(B_T-V_T)$ & Tycho & 1.02 & \nodata & \nodata \\
$(V_T-K_2)$ & Tycho-2MASS & 2.87 & \nodata & \nodata
\enddata
\tablenotetext{1}{$k=E(\mathrm{color})/E(B-V)$.}
\label{t:red}
\end{deluxetable*}

\section{The calibrations} \label{sect:fits}

The fits to the data were performed in a two-step procedure described as follows:

(1) All data points were iteratively fitted to
\begin{eqnarray}
\theta_\mathrm{eff} & = & a_0+a_1X+a_2X^2+a_3X\feh \nonumber \\
& & +a_4\feh+a_5\feh^2\ ,\label{eq:theta}
\end{eqnarray}
where $\theta_\mathrm{eff}=5040/\teff$, $X$ represents the color, and $a_i$ ($i=1,\ldots,5$) the coefficients of the fit. In every iteration, the points departing more than $2.5\sigma$ from the mean fit were discarded. Normally, five to seven iterations were required.

The particular analytical expresion adopted (Eq.~\ref{eq:theta}) reasonably reproduces the observed trends, and it has also some physical meaning (see e.g. Sect. 3 in RM04a).

(2) Whenever necessary and reasonable (see below), the residuals of the fit ($\teff^\mathrm{IRFM}-\teff^\mathrm{cal}$) were fitted to polynomials in $X$ to remove any small systematic trends due to the incapability of Eq. (\ref{eq:theta}) to reproduce the effects of spectral features such as the Balmer lines, the G band, or the Paschen Jump on the observed colors. The polynomial fits were performed in metallicity bins, and since they rarely exceed 50~K, continuity is not severely compromised. The polynomial fits $P(X,\feh)$ need to be added to Eq.~(\ref{eq:theta}) so the final form is
\begin{equation}
\teff=\frac{5040}{\theta_\mathrm{eff}}+P(X,\feh)\ .\label{eq:teff}
\end{equation}

Polynomial fits were only performed when enough stars defined a clear trend in the residuals and care was taken as not to force unphysical, artificial results. Neglecting the polynomial fits in this procedure would lead to systematic errors of the order of 30~K or 40~K.

Due to the nature of the fits (particularly the polynomial corrections), extrapolation leads to unreliable results. If necessary, one may, as a last resort, extrapolate from the tables given in \S4. 

\begin{figure*}
\epsscale{1.}
\plotone{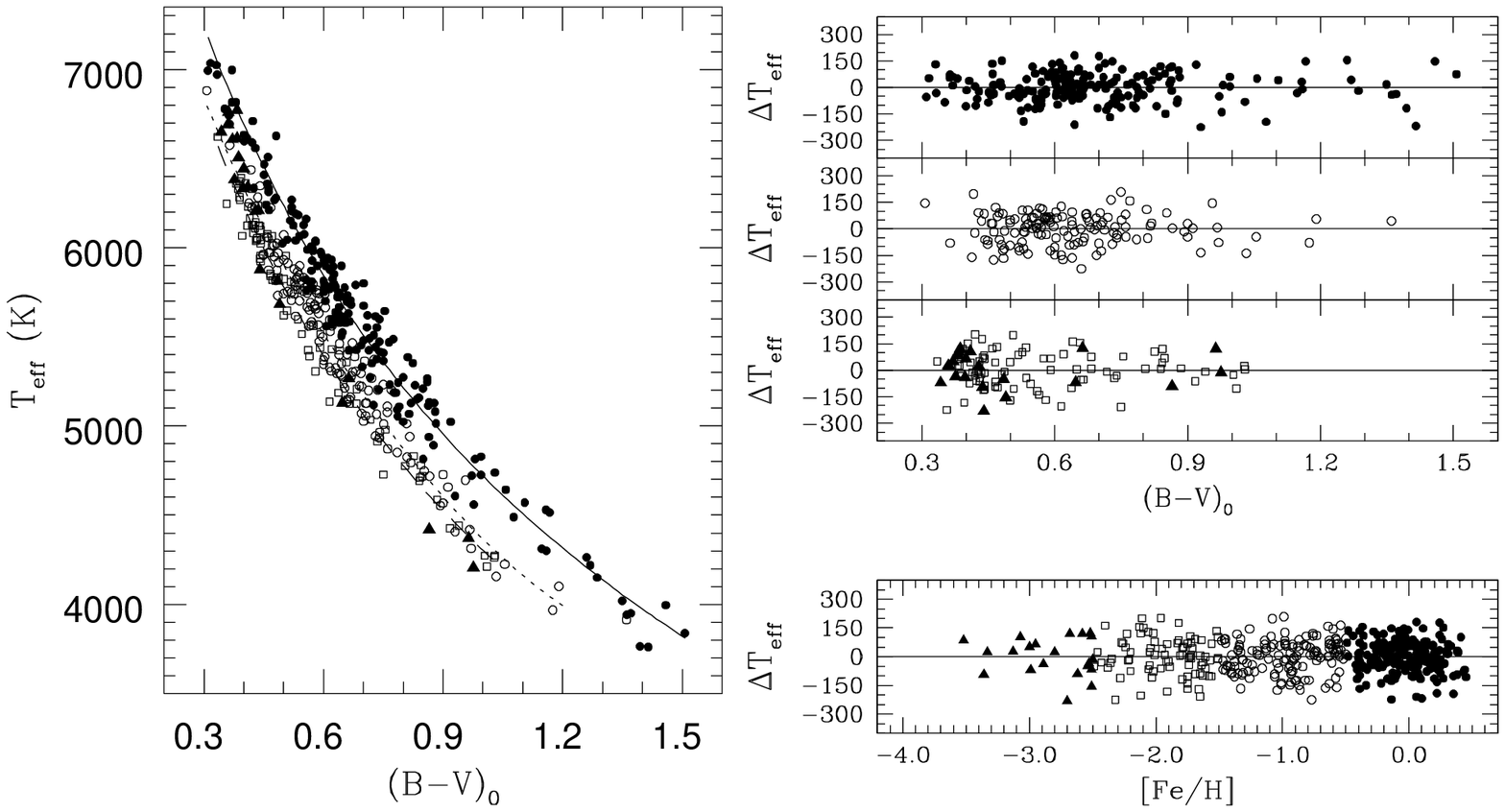} \caption{Left: $\teff$ vs. $(B-V)$ observed for dwarfs in the metallicity bins: $-0.5<\feh\leq+0.5$ (filled circles), $-1.5<\feh\leq-0.5$ (open circles), $-2.5<\feh\leq-1.5$ (squares), and $\feh\leq-2.5$ (triangles). The lines corresponding to our calibration for $\feh=0.0$ (solid line), $\feh=-1.0$ (dotted line), and $\feh=-2.0$ (dashed line) are also shown. Right: residuals of the fit ($\Delta\teff=\teff^\mathrm{IRFM}-\teff^\mathrm{cal}$) as a function of color and $\feh$.} \label{fig:dbv}
\end{figure*}

\begin{figure*}
\epsscale{1.}
\plotone{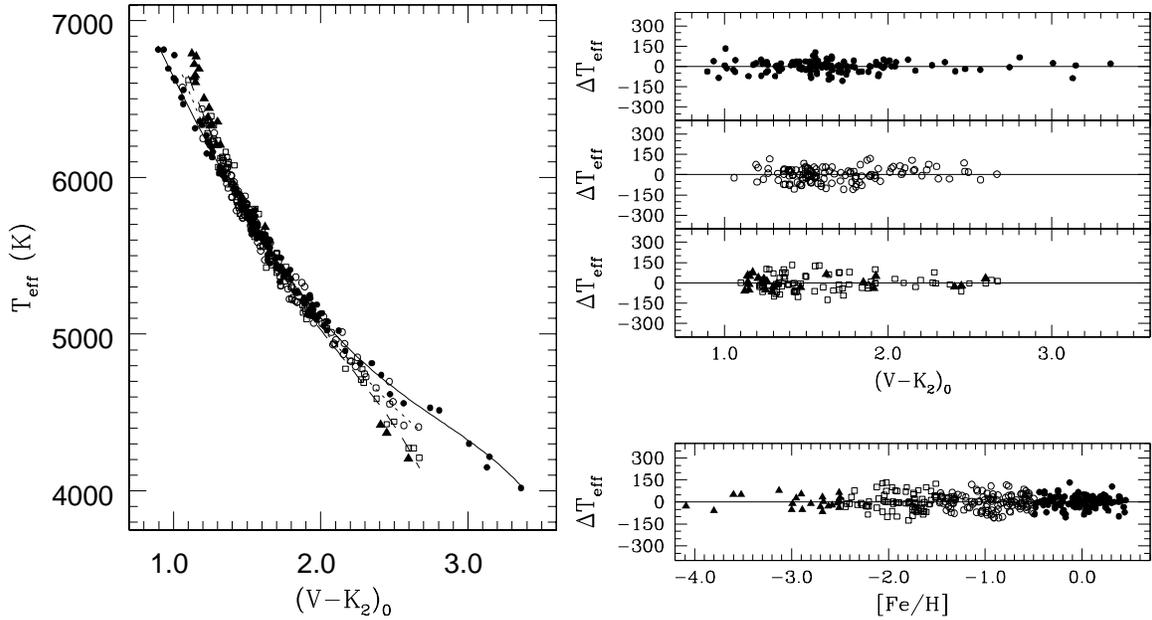} \caption{As in Fig. \ref{fig:dbv} for $(V-K_2)$ (dwarfs).} \label{fig:dvk2}
\end{figure*}

\begin{figure*}
\epsscale{1.}
\plotone{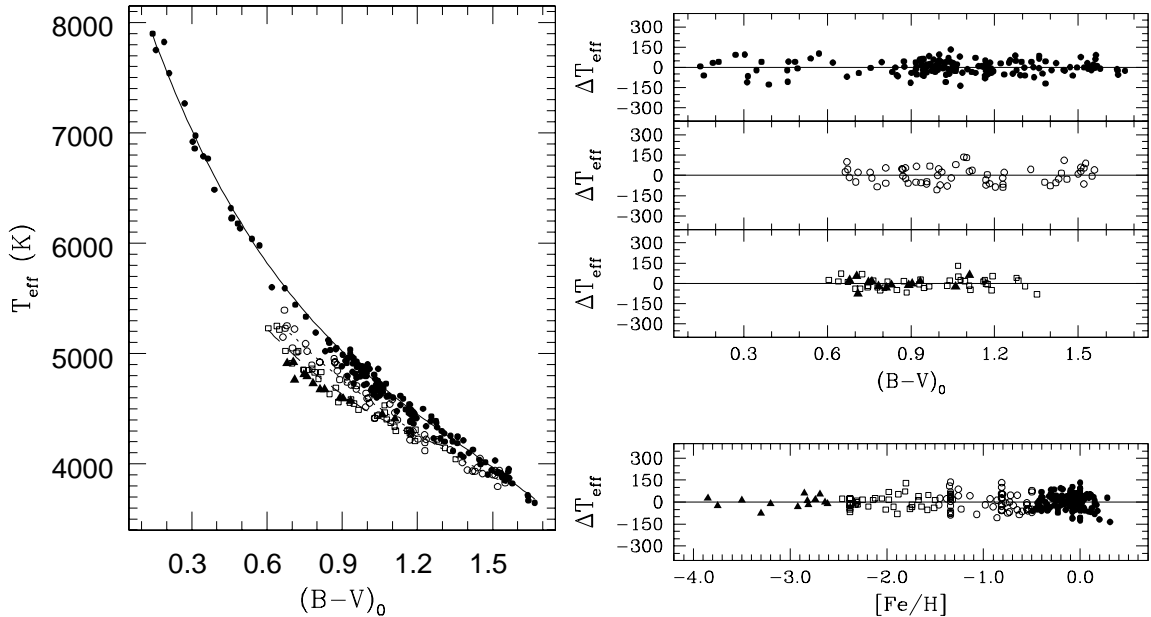} \caption{As in Fig. \ref{fig:dbv} for $(B-V)$ (giants).} \label{fig:gbv}
\end{figure*}

\begin{figure*}
\epsscale{1.}
\plotone{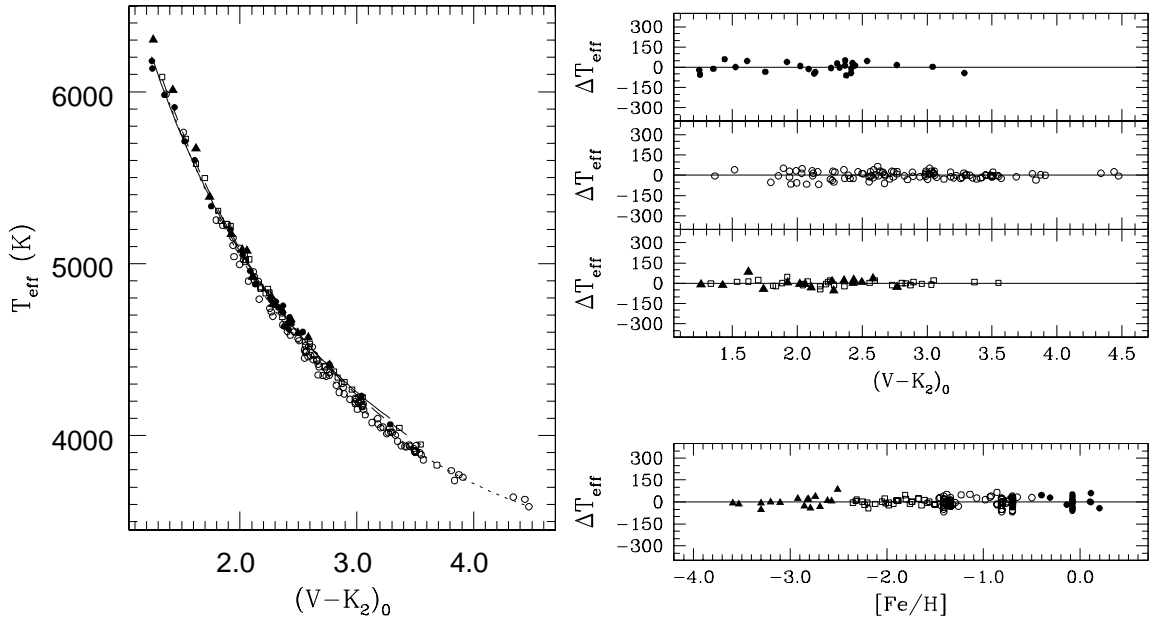} \caption{As in Fig. \ref{fig:dbv} for $(V-K_2)$ (giants).} \label{fig:gvk2}
\end{figure*}

The coefficients of the fits ($a_i$), the number of stars included ($N$), and the standard deviations ($\sigma(\teff)$) of the seventeen color calibrations performed are given in Tables \ref{t:aD} (dwarfs) and \ref{t:aG} (giants). The ranges of applicability ($X_\mathrm{min}<X<X_\mathrm{max}$) in the following metallicity bins: $-0.5<\feh<+0.5$, $-1.5<\feh\leq-0.5$, $-2.5<\feh\leq-1.5$, $-3.5<\feh\leq-2.5$; and the coefficients of the polynomial fits ($P=\sum_i P_i X^i$) are given in Tables \ref{t:pD} (dwarfs) and \ref{t:pG} (giants). Figs. \ref{fig:dbv}-\ref{fig:gvk2} illustrate some of the calibrations in the $\teff$ vs. color planes, and the residuals of the fits (after the polynomial corrections). The complete set of figures, for all the calibrations, is available online.\footnote{\ https://webspace.utexas.edu/ir68/teff} The figures illustrate, better than the tables show, how far in $\feh$ (both at the metal-poor and metal-rich ends) a particular color calibration may be applied.

\begin{deluxetable*}{rrrrrrrrr}
\tablecaption{Coefficients of the Dwarf Star Color Calibrations\tablenotemark{1}}
\tablehead{\colhead{color ($X$)} & \colhead{$a_0$} & \colhead{$a_1$} & \colhead{$a_2$} & \colhead{$a_3$} & \colhead{$a_4$} & \colhead{$a_5$} & \colhead{$N$} & \colhead{$\sigma(\teff)$}}
\startdata
$(B-V)$ & 0.5002 & 0.6440 & -0.0690 & -0.0230 & -0.0566 & -0.0170 & 495 & 88 \\
$(b-y)$ & 0.4129 & 1.2570 & -0.2268 & -0.0242 & -0.0464 & -0.0200 & 434 & 87 \\
$(Y-V)$ & 0.0644 & 1.7517 & -0.5264 & -0.0044 & -0.0407 & -0.0132 & 159 & 121 \\
$(V-S)$ & 0.2417 & 1.3653 & -0.3823 & -0.0387 & -0.0105 & -0.0077 & 142 & 95 \\
$(B_2-V_1)$ & 0.6019 & 0.7663 & -0.0713 & -0.0339 & -0.0382 & -0.0137 & 358 & 74 \\
$(B_2-G)$ & 0.8399 & 0.4909 & -0.0666 & -0.0360 & -0.0468 & -0.0124 & 368 & 66 \\
$t$ & 0.7696 & 0.5927 & 0.3439 & -0.0437 & -0.0143 & -0.0088 & 308 & 66 \\
$(V-R_C)$ & 0.4333 & 1.4399 & -0.5419 & -0.0481 & -0.0239 & -0.0125 & 133 & 84 \\
$(V-I_C)$ & 0.3295 & 0.9516 & -0.2290 & -0.0316 &  0.0003 & -0.0081 & 127 & 68 \\
$(R_C-I_C)$ & 0.2919 & 2.1141 & -1.0723 & -0.0756 &  0.0267 & -0.0041 & 137 & 76 \\
$C(42-45)$ & 0.5153 & 0.5963 & -0.0572 & -0.0573 & -0.0221 & -0.0018 & 120 & 70 \\
$C(42-48)$ & 0.1601 & 0.4533 & -0.0135 & -0.0471 &  0.0305 & -0.0020 & 133 & 70 \\
$(B_T-V_T)$ & 0.5619 & 0.4462 & -0.0029 & 0.0003 & -0.0746 & -0.0190 & 378 & 104 \\
$(V-J_2)$ & 0.4050 & 0.4792 & -0.0617 & -0.0392 & 0.0401 & -0.0023 & 361 & 62 \\
$(V-H_2)$ & 0.4931 & 0.3056 & -0.0241 & -0.0396 & 0.0678 & 0.0020 & 364 & 57 \\
$(V-K_2)$ & 0.4942 & 0.2809 & -0.0180 & -0.0294 & 0.0444 & -0.0008 & 397 & 50 \\
$(V_T-K_2)$ & 0.4886 & 0.2773 & -0.0195 & -0.0300 & 0.0467 & -0.0008 & 318 & 59 
\enddata
\tablenotetext{1}{$N$ is the number of stars employed and $\sigma$ the standard deviation of each fit.}
\label{t:aD}
\end{deluxetable*}

\begin{deluxetable*}{rrrrrrrrr}
\tablecaption{Coefficients of the Giant Star Color Calibrations\tablenotemark{1}}
\tablehead{\colhead{color ($X$)} & \colhead{$a_0$} & \colhead{$a_1$} & \colhead{$a_2$} & \colhead{$a_3$} & \colhead{$a_4$} & \colhead{$a_5$} & \colhead{$N$} & \colhead{$\sigma(\teff)$}}
\startdata
$(B-V)$ &  0.5737 & 0.4882 & -0.0149 & 0.0563 & -0.1160 & -0.0114 & 269 & 51 \\
$(b-y)$ & 0.5515 & 0.9085 & -0.1494 & 0.0616 & -0.0668 & -0.0083 & 208 & 68 \\
$(Y-V)$ & 0.3672 & 1.0467 & -0.1995 & 0.0650 & -0.0913 & -0.0133 & 159 & 78 \\
$(V-S)$ & 0.3481 & 1.1188 & -0.2068 & 0.0299 & -0.0481 & -0.0083 & 152 & 69 \\
$(B_2-V_1)$ & 0.6553 & 0.6278 & -0.0629 & 0.0627 & -0.0816 & -0.0084 & 200 & 45 \\
$(B_2-G)$ & 0.8492 & 0.4344 & -0.0365 & 0.0466 & -0.0696 & -0.0107 & 189 & 39 \\
$t$ & 0.7460 & 0.8151 & -0.1943 & 0.0855 & -0.0421 & -0.0034 & 192 & 44 \\
$(V-R_C)$ & 0.3849 & 1.6205 & -0.6395 & 0.1060 & -0.0875 & -0.0089 & 90 & 41 \\
$(V-I_C)$ & 0.3575 & 0.9069 & -0.2025 & 0.0395 & -0.0551 & -0.0061 & 95 & 40 \\
$(R_C-I_C)$ & 0.4351 & 1.6549 & -0.7215 & -0.0610 & 0.0332 & -0.0023 & 128 & 62 \\
$C(42-45)$ & 0.4783 & 0.7748 & -0.1361 & -0.0712 & -0.0117 & 0.0071 & 188 & 57 \\
$C(42-48)$ & 0.0023 & 0.6401 & -0.0632 & -0.0023 & -0.0706 & -0.0070 & 191 & 49 \\
$(B_T-V_T)$ & 0.5726 & 0.4461 & -0.0324 & 0.0518 & -0.1170 & -0.0094 & 261 & 82 \\
$(V-J_2)$ & 0.2943 & 0.5604 & -0.0677 & 0.0179 & -0.0532 & -0.0088 & 163 & 38 \\
$(V-H_2)$ & 0.4354 & 0.3405 & -0.0263 & -0.0012 & -0.0049 & -0.0027 & 177 & 32 \\
$(V-K_2)$ & 0.4405 & 0.3272 & -0.0252 & -0.0016 & -0.0053 & -0.0040 & 182 & 28 \\
$(V_T-K_2)$ & 0.4813 & 0.2871 & -0.0203 & -0.0045 & 0.0062 & -0.0019 & 112 & 39 
\enddata
\tablenotetext{1}{$N$ is the number of stars employed and $\sigma$ the standard deviation of each fit.}
\label{t:aG}
\end{deluxetable*}

\begin{deluxetable*}{rrrrrrrrrrrr}
\tabletypesize{\scriptsize}
\tablecaption{Ranges of Applicability per Metallicity Bin and Coefficients of the Polynomial Fits for the Dwarf Star Calibrations}
\tablehead{\colhead{color ($X$)} & $\feh$\tablenotemark{1} & \colhead{$X_\mathrm{min}$} & \colhead{$X_\mathrm{max}$} & \colhead{$P_0$} & \colhead{$P_1$} & \colhead{$P_2$} & \colhead{$P_3$} & \colhead{$P_4$} & \colhead{$P_5$} & \colhead{$P_6$} }
\startdata
$(B-V)$ & $+0.0$ & 0.310 & 1.507 & -261.548 & 684.977 & -470.049 & 79.8977 & \nodata & \nodata & \nodata\\
$(B-V)$ & $-1.0$ & 0.307 & 1.202 & -324.033 & 1516.44 & -2107.37 & 852.150 & \nodata & \nodata & \nodata\\
$(B-V)$ & $-2.0$ & 0.335 & 1.030 & 30.5985 & -46.7882 & \nodata & \nodata & \nodata & \nodata & \nodata\\
$(B-V)$ & $-3.0$ & 0.343 & 0.976 & 139.965 & -292.329 & \nodata & \nodata & \nodata & \nodata & \nodata\\ \hline
$(b-y)$ & $+0.0$ & 0.248 & 0.824 & -1237.11 &  6591.29 & -11061.3 & 5852.18 & \nodata & \nodata & \nodata\\
$(b-y)$ & $-1.0$ & 0.234 & 0.692 & -2617.66 &  22607.4 & -68325.4 & 86072.5 & -38602.2 & \nodata & \nodata\\
$(b-y)$ & $-2.0$ & 0.290 & 0.672 &  103.927 & -312.419 &  225.430 & \nodata & \nodata & \nodata & \nodata\\
$(b-y)$ & $-3.0$ & 0.270 & 0.479 & -294.106 &  648.320 & \nodata & \nodata & \nodata & \nodata & \nodata\\ \hline
$(Y-V)$ & $+0.0$ & 0.420 & 0.940 & -10407.1 & 42733.6 & -27378.8 & -96466.3 & 162033. & -70956.4 &
\nodata\\
$(Y-V)$ & $-1.0$ & 0.452 & 0.660 &  11.6451 & \nodata & \nodata & \nodata & \nodata & \nodata & \nodata\\
$(Y-V)$ & $-2.0$ & 0.455 & 0.720 & -507.732 & 1943.73 & -1727.66 & \nodata & \nodata & \nodata & \nodata\\
$(Y-V)$ & $-3.0$ & 0.446 & 0.643 & -310.166 & 496.709 & \nodata & \nodata & \nodata & \nodata & \nodata\\ \hline
$(V-S)$ & $+0.0$ & 0.370 & 1.130 & -1436.48 & 5566.00 & -6780.53 & 2613.40 & \nodata & \nodata & \nodata\\
$(V-S)$ & $-1.0$ & 0.410 & 0.690 & -728.818 & 2256.18 & -1704.54 & \nodata & \nodata & \nodata & \nodata\\
$(V-S)$ & $-2.0$ & 0.441 & 0.810 & 101.031 & 114.354 & -447.778 & \nodata & \nodata & \nodata & \nodata\\
$(V-S)$ & $-3.0$ & 0.438 & 0.584 & 596.461 & -1130.92 & \nodata & \nodata & \nodata & \nodata & \nodata\\ \hline
$(B_2-V_1)$ & $+0.0$ & 0.119 & 0.936 & -439.817 & 2637.06 & -4762.80 & 2606.79 & \nodata & \nodata & \nodata \\
$(B_2-V_1)$ & $-1.0$ & 0.132 & 0.593 & -257.527 & 2078.96 & -4919.04 & 3685.65 & -500.348 & \nodata &
\nodata\\
$(B_2-V_1)$ & $-2.0$ & 0.178 & 0.621 & -28.5544 & 228.735 & -295.958 & \nodata & \nodata & \nodata & \nodata\\
$(B_2-V_1)$ & $-3.0$ & 0.185 & 0.435 &  64.2911 & -365.124 & \nodata & \nodata & \nodata & \nodata & \nodata\\ \hline
$(B_2-G)$ & $+0.0$ & -0.271 & 1.110 & -6.29800 &  160.976 & -386.520 & 250.628 & \nodata & \nodata & \nodata\\
$(B_2-G)$ & $-1.0$ & -0.262 & 0.502 &  21.3254 & -56.4562 & -651.533 & 720.639 & \nodata & \nodata & \nodata\\
$(B_2-G)$ & $-2.0$ & -0.200 & 0.544 &  11.5114 & -34.5752 & -265.563 & \nodata & \nodata & \nodata & \nodata\\
$(B_2-G)$ & $-3.0$ & -0.179 & 0.150 & -156.547 & -313.408 &  4886.53 & \nodata & \nodata & \nodata & \nodata\\ \hline
$t$ & $+0.0$ & -0.119 & 0.450 & -16.3530 & 273.725 & -1383.02 & 2274.81 & \nodata & \nodata & \nodata\\
$t$ & $-1.0$ & -0.066 & 0.373 & 35.2419 & -185.953 & \nodata & \nodata & \nodata & \nodata & \nodata\\
$t$ & $-2.0$ & -0.006 & 0.333 & 11.9635 & \nodata & \nodata & \nodata & \nodata & \nodata & \nodata\\
$t$ & $-3.0$ &  0.020 & 0.295 & -39.1918 & \nodata & \nodata & \nodata & \nodata & \nodata & \nodata\\ \hline
$(V-R_C)$ & $+0.0$ & 0.204 & 0.880 & -2666.55 & 27264.5 & -103923. & 174663. & -104940. & -23249.4 & 32644.9 \\
$(V-R_C)$ & $-1.0$ & 0.284 & 0.546 & 4.20153 & \nodata & \nodata & \nodata & \nodata & \nodata & \nodata\\
$(V-R_C)$ & $-2.0$ & 0.264 & 0.532 & 123.940 & -342.217 & \nodata & \nodata & \nodata & \nodata & \nodata\\
$(V-R_C)$ & $-3.0$ & 0.240 & 0.336 & 8.55498 & \nodata & \nodata & \nodata & \nodata & \nodata & \nodata\\ \hline

$(V-I_C)$ & $+0.0$ & 0.491 & 1.721 & -2757.79 & 9961.33 & -10546.6 & -1746.05 & 10512.3 & -6653.57 & 1301.21\\
$(V-I_C)$ & $-1.0$ & 0.597 & 1.052 & -22.9008 & 40.2078 & \nodata & \nodata & \nodata & \nodata & \nodata\\
$(V-I_C)$ & $-2.0$ & 0.547 & 1.026 & -667.732 & 1709.88 & -1069.62 & \nodata & \nodata & \nodata & \nodata\\ \hline
$(R_C-I_C)$ & $+0.0$ & 0.242 & 0.838 & -3326.97 & 26263.8 & -75355.8 & 94246.5 & -43334.8 & \nodata & \nodata\\
$(R_C-I_C)$ & $-1.0$ & 0.300 & 0.718 & 12.4740 & \nodata & \nodata & \nodata & \nodata & \nodata & \nodata\\
$(R_C-I_C)$ & $-2.0$ & 0.283 & 0.551 & -5837.31 & 41439.2 & -94729.8 & 69584.8 & \nodata & \nodata & \nodata\\
$(R_C-I_C)$ & $-3.0$ & 0.290 & 0.364 & 32.1826 & \nodata & \nodata & \nodata & \nodata & \nodata & \nodata\\ \hline
$C(42-45)$ & $+0.0$ & 0.461 & 1.428 & 1533.40 & -5546.94 & 6324.29 & -2254.52 & \nodata & \nodata & \nodata\\
$C(42-45)$ & $-1.0$ & 0.480 & 0.812 & 808.065 & -2725.54 & 2806.13 & -902.995 & \nodata & \nodata & \nodata\\ \hline
$C(42-48)$ & $+0.0$ & 1.286 & 2.711 & 658.568 & -283.310 & -709.877 & 575.693 & -114.834 & \nodata & \nodata\\
$C(42-48)$ & $-1.0$ & 1.465 & 1.957 & 176.678 & -204.699 & 53.2421 & \nodata & \nodata & \nodata & \nodata\\
$C(42-48)$ & $-2.0$ & 1.399 & 1.509 & 1069.18 & -678.907 & \nodata & \nodata & \nodata & \nodata & \nodata\\ \hline
$(B_T-V_T)$ & $+0.0$ & 0.344 & 1.715 & 1199.21 & -5470.57 & 8367.46 & -5119.55 & 1078.09 & \nodata & \nodata\\
$(B_T-V_T)$ & $-1.0$ & 0.391 & 1.556 & -64.1045 & 140.575 & -59.4233 & \nodata & \nodata & \nodata & \nodata\\
$(B_T-V_T)$ & $-2.0$ & 0.380 & 0.922 & -6030.19 & 29153.4 & -25882.7 & -64112.9 & 126115. & -59817.9 & \nodata\\
$(B_T-V_T)$ & $-3.0$ & 0.367 & 0.504 & -3255.07 & 16259.9 & -20315.3 & \nodata & \nodata & \nodata & \nodata\\ \hline
$(V-J_2)$ & $+0.0$ & 0.815 & 2.608 &  422.406 & -910.603 & 621.335  & -132.566 & \nodata & \nodata & \nodata\\
$(V-J_2)$ & $-1.0$ & 0.860 & 2.087 & -466.616 & 658.349  & -220.454 & \nodata & \nodata & \nodata & \nodata\\
$(V-J_2)$ & $-2.0$ & 0.927 & 1.983 & -862.072 & 1236.84  & -423.729 & \nodata & \nodata & \nodata & \nodata\\
$(V-J_2)$ & $-3.0$ & 0.891 & 1.932 & -1046.10 & 1652.06  & -597.340 & \nodata & \nodata & \nodata & \nodata\\ \hline
$(V-H_2)$ & $+0.0$ & 0.839 & 3.215 & -53.5574 & 36.0990 & 15.6878 & -8.84468 & \nodata & \nodata & \nodata\\
$(V-H_2)$ & $-1.0$ & 1.032 & 2.532 & 1.60629 & \nodata & \nodata & \nodata & \nodata & \nodata & \nodata\\
$(V-H_2)$ & $-2.0$ & 1.070 & 2.535 & 506.559 & -1277.52 & 939.519 & -208.621 & \nodata & \nodata & \nodata\\
$(V-H_2)$ & $-3.0$ & 1.093 & 2.388 & -471.588 & 643.972 & -199.639 & \nodata & \nodata & \nodata & \nodata\\ \hline
$(V-K_2)$ & $+0.0$ & 0.896 & 3.360 & -1425.36 & 3218.36 & -2566.54 & 859.644 & -102.554 & \nodata & \nodata\\
$(V-K_2)$ & $-1.0$ & 1.060 & 2.665 & 2.35133 & \nodata & \nodata & \nodata & \nodata & \nodata & \nodata\\
$(V-K_2)$ & $-2.0$ & 1.101 & 2.670 & -1849.46 & 4577.00 & -4284.02 & 1770.38 & -268.589 & \nodata & \nodata\\
$(V-K_2)$ & $-3.0$ & 1.126 & 2.596 & 215.721 & -796.519 & 714.423 & -175.678 & \nodata & \nodata & \nodata\\ \hline
$(V_T-K_2)$ & $+0.0$ & 0.942 & 3.284 & -1581.85 & 3273.10 & -2395.38 & 736.352 & -80.8177 & \nodata & \nodata\\
$(V_T-K_2)$ & $-1.0$ & 1.078 & 2.561 &  68.1279 & -130.968 & 52.8391 & \nodata & \nodata & \nodata & \nodata\\
$(V_T-K_2)$ & $-2.0$ & 1.237 & 2.406 & -2384.82 & 4196.14 & -2557.04 & 595.365 & -31.9955 & \nodata & \nodata\\
$(V_T-K_2)$ & $-3.0$ & 1.170 & 1.668 & -628.682 & 423.682 & \nodata & \nodata & \nodata & \nodata & \nodata 
\enddata
\tablenotetext{1}{Metallicity bins coded as: $\feh=+0.0$ ($-0.5<\feh<+0.5$), $\feh=-1.0$ ($-1.5<\feh\leq-0.5$), $\feh=-2.0$ ($-2.5<\feh\leq-1.5$), $\feh=-3.0$ ($-4.0<\feh\leq-2.5$).}
\label{t:pD}
\end{deluxetable*}

\begin{deluxetable*}{rrrrrrrrrrrr}
\tabletypesize{\scriptsize}
\tablecaption{Ranges of Applicability per Metallicity Bin and Coefficients of the Polynomial Fits for the Giant Star Calibrations}
\tablehead{\colhead{color ($X$)} & $\feh$\tablenotemark{1} & \colhead{$X_\mathrm{min}$} & \colhead{$X_\mathrm{max}$} & \colhead{$P_0$} & \colhead{$P_1$} & \colhead{$P_2$} & \colhead{$P_3$} & \colhead{$P_4$} & \colhead{$P_5$} & \colhead{$P_6$} }
\startdata
$(B-V)$ & $+0.0$ & 0.144 & 1.668 & 112.116 & -372.622 & 67.1254 & 395.333 & -203.471 & \nodata & \nodata\\
$(B-V)$ & $-1.0$ & 0.664 & 1.558 & -12.9762 & \nodata & \nodata & \nodata & \nodata & \nodata & \nodata\\
$(B-V)$ & $-2.0$ & 0.605 & 1.352 & 606.032 & -1248.79 & 627.453 & \nodata & \nodata & \nodata & \nodata\\
$(B-V)$ & $-3.0$ & 0.680 & 1.110 & -9.26209 & \nodata & \nodata & \nodata & \nodata & \nodata & \nodata\\ \hline
$(b-y)$ & $+0.0$ & 0.053 & 1.077 & -124.159 & 553.827 & -490.703 & \nodata & \nodata & \nodata & \nodata\\
$(b-y)$ & $-1.0$ & 0.309 & 0.893 & 888.088 & -2879.23 & 2097.89 & \nodata & \nodata & \nodata & \nodata\\
$(b-y)$ & $-2.0$ & 0.388 & 0.702 & 1867.63 & -6657.49 & 5784.81 & \nodata & \nodata & \nodata & \nodata\\
$(b-y)$ & $-3.0$ & 0.404 & 0.683 & 348.237 & -659.093 & \nodata & \nodata & \nodata & \nodata & \nodata\\ \hline
$(Y-V)$ & $+0.0$ & 0.230 & 1.290 & -308.851 & 1241.57 & -1524.60 & 593.157 & \nodata & \nodata & \nodata\\
$(Y-V)$ & $-1.0$ & 0.558 & 0.940 & -36.6533 & 383.901 & -458.085 & \nodata & \nodata & \nodata & \nodata\\
$(Y-V)$ & $-2.0$ & 0.544 & 0.817 & 3038.83 & -8668.15 & 6067.04 & \nodata & \nodata & \nodata & \nodata\\
$(Y-V)$ & $-3.0$ & 0.510 & 0.830 & 2685.88 & -7433.07 & 4991.81 & \nodata & \nodata & \nodata & \nodata\\ \hline
$(V-S)$ & $+0.0$ & 0.261 & 1.230 & -1605.54 & 9118.16 & -17672.6 & 14184.1 & -4023.76 & \nodata & \nodata\\
$(V-S)$ & $-1.0$ & 0.508 & 0.992 & 187.841 & -270.092 & \nodata & \nodata & \nodata & \nodata & \nodata\\
$(V-S)$ & $-2.0$ & 0.529 & 0.990 & 10.1750 & \nodata & \nodata & \nodata & \nodata & \nodata & \nodata\\
$(V-S)$ & $-3.0$ & 0.573 & 0.790 & -14.2019 & \nodata & \nodata & \nodata & \nodata & \nodata & \nodata\\ \hline
$(B_2-V_1)$ & $+0.0$ & -0.079 & 1.321 & -15.0383 & 50.8876 &  -32.3978 & \nodata & \nodata & \nodata & \nodata\\
$(B_2-V_1)$ & $-1.0$ & 0.385 & 1.021 & 80.1344 & -147.055 & \nodata & \nodata & \nodata & \nodata & \nodata\\
$(B_2-V_1)$ & $-2.0$ & 0.307 & 0.958 & 323.889 & -1031.06 & 795.024 & \nodata & \nodata & \nodata & \nodata\\
$(B_2-V_1)$ & $-3.0$ & 0.407 & 0.648 & 1403.86 & -4866.09 & 4029.75 & \nodata & \nodata & \nodata & \nodata\\ \hline
$(B_2-G)$ & $+0.0$ & -0.543 & 1.230 & -0.52642 & 10.4471 & -7.53155 & \nodata & \nodata & \nodata & \nodata\\
$(B_2-G)$ & $-1.0$ & 0.155 & 0.966 & 26.1904 & -89.2171 & \nodata & \nodata & \nodata & \nodata & \nodata\\
$(B_2-G)$ & $-2.0$ & 0.132 & 0.991 & 9.87980 & \nodata & \nodata & \nodata & \nodata & \nodata & \nodata\\
$(B_2-G)$ & $-3.0$ & 0.104 & 0.437 & 232.248 & -1452.43 & 1848.07 & \nodata & \nodata & \nodata & \nodata\\ \hline
$t$ & $+0.0$ & 0.072 & 0.970 & -46.1506 & -60.1641 & 643.522 & -599.555 & \nodata & \nodata & \nodata & \nodata\\
$t$ & $-1.0$ & 0.064 & 0.766 & 27.8739 & -84.1166 & \nodata & \nodata & \nodata & \nodata & \nodata & \nodata\\
$t$ & $-2.0$ & 0.166 & 0.619 & 67.1191 & -139.127 & \nodata & \nodata & \nodata & \nodata & \nodata & \nodata\\
$t$ & $-3.0$ & 0.215 & 0.511 & 122.254 & -394.604 & \nodata & \nodata & \nodata & \nodata & \nodata & \nodata\\ \hline
$(V-R_C)$ & $+0.0$ & 0.299 & 1.106 & -8.51797 & 15.6675 & \nodata & \nodata & \nodata & \nodata & \nodata\\
$(V-R_C)$ & $-1.0$ & 0.387 & 0.752 & -10.7764 & \nodata & \nodata & \nodata & \nodata & \nodata & \nodata\\
$(V-R_C)$ & $-2.0$ & 0.429 & 0.598 & 61.9821 & -78.7382 & \nodata & \nodata & \nodata & \nodata & \nodata\\
$(V-R_C)$ & $-3.0$ & 0.394 & 0.550 & 27.9886 & -100.149 & \nodata & \nodata & \nodata & \nodata & \nodata\\ \hline
$(V-I_C)$ & $+0.0$ & 0.573 & 2.000 & 0.42933 & \nodata & \nodata & \nodata & \nodata & \nodata & \nodata\\
$(V-I_C)$ & $-1.0$ & 0.795 & 1.524 & -0.14180 & \nodata & \nodata & \nodata & \nodata & \nodata & \nodata\\
$(V-I_C)$ & $-2.0$ & 0.870 & 1.303 & 9.31011 & \nodata & \nodata & \nodata & \nodata & \nodata & \nodata\\
$(V-I_C)$ & $-3.0$ & 0.812 & 1.095 & -23.0514 & \nodata & \nodata & \nodata & \nodata & \nodata & \nodata\\ \hline
$(R_C-I_C)$ & $+0.0$ & 0.413 & 0.793 & 61.3557 & -116.711 & \nodata & \nodata & \nodata & \nodata & \nodata\\
$(R_C-I_C)$ & $-1.0$ & 0.383 & 0.771 & -16.8645 & \nodata & \nodata & \nodata & \nodata & \nodata & \nodata\\
$(R_C-I_C)$ & $-2.0$ & 0.434 & 0.725 & 32.0870 & \nodata & \nodata & \nodata & \nodata & \nodata & \nodata\\
$(R_C-I_C)$ & $-3.0$ & 0.364 & 0.545 & -15.6318 & \nodata & \nodata & \nodata & \nodata & \nodata & \nodata\\ \hline
$C(42-45)$ & $+0.0$ & 0.409 & 1.369 & -68.3798 & 109.259 & -34.4503 & \nodata & \nodata & \nodata & \nodata\\
$C(42-45)$ & $-1.0$ & 0.430 & 1.270 & -0.62507 & \nodata & \nodata & \nodata & \nodata & \nodata & \nodata\\
$C(42-45)$ & $-2.0$ & 0.441 & 0.894 & -40.0150 & 35.6803 & \nodata & \nodata & \nodata & \nodata & \nodata\\
$C(42-45)$ & $-3.0$ & 0.490 & 0.640 & -314.177 & 636.443 & \nodata & \nodata & \nodata & \nodata & \nodata\\ \hline
$C(42-48)$ & $+0.0$ & 1.531 &  2.767 & 1006.40 & -549.012 & -649.212 & 534.912 & -100.038 & \nodata & \nodata\\
$C(42-48)$ & $-1.0$ & 1.400 &  2.647 & -6.92065 & \nodata & \nodata & \nodata & \nodata & \nodata & \nodata\\
$C(42-48)$ & $-2.0$ & 1.466 &  2.260 & -113.222 & 57.3030 & \nodata & \nodata & \nodata & \nodata & \nodata\\
$C(42-48)$ & $-3.0$ & 1.571 &  1.799 & 566.914 & -329.631 & \nodata & \nodata & \nodata & \nodata & \nodata\\ \hline
$(B_T-V_T)$ & $+0.0$ & 0.123 & 1.953 & 346.881 & -1690.16 & 2035.65 & -797.248 & 70.7799 & \nodata & \nodata\\
$(B_T-V_T)$ & $-1.0$ & 0.424 & 1.644 & 196.416 & -372.164 & 126.196 & \nodata & \nodata & \nodata & \nodata\\
$(B_T-V_T)$ & $-2.0$ & 0.534 & 1.356 & 938.789 & -1919.98 & 929.779 & \nodata & \nodata & \nodata & \nodata\\
$(B_T-V_T)$ & $-3.0$ & 0.465 & 1.026 & 1112.46 & -2717.81 & 1577.18 & \nodata & \nodata & \nodata & \nodata\\ \hline
$(V-J_2)$ & $+0.0$ & 1.259 & 2.400 & -122.595 & 76.4847 & \nodata & \nodata & \nodata & \nodata & \nodata\\
$(V-J_2)$ & $-1.0$ & 1.030 & 3.418 & -10.3848 & \nodata & \nodata & \nodata & \nodata & \nodata & \nodata\\
$(V-J_2)$ & $-2.0$ & 1.033 & 2.679 & 4.18695 & 13.8937 & \nodata & \nodata & \nodata & \nodata & \nodata\\
$(V-J_2)$ & $-3.0$ & 0.977 & 2.048 & -67.7716 & 28.9202 & \nodata & \nodata & \nodata & \nodata & \nodata\\ \hline
$(V-H_2)$ & $+0.0$ & 1.194 & 3.059 & -377.022 & 334.733 & -69.8093 & \nodata & \nodata & \nodata & \nodata\\
$(V-H_2)$ & $-1.0$ & 1.293 & 4.263 & 71.7949 & -55.5383 & 9.61821 & \nodata & \nodata & \nodata & \nodata\\
$(V-H_2)$ & $-2.0$ & 1.273 & 3.416 & -27.4190 & 20.7082 & \nodata & \nodata & \nodata & \nodata & \nodata\\
$(V-H_2)$ & $-3.0$ & 1.232 & 2.625 & -46.2946 & 20.1061 & \nodata & \nodata & \nodata & \nodata & \nodata\\ \hline
$(V-K_2)$ & $+0.0$ & 1.244 & 3.286 & -72.6664 & 36.5361 & \nodata & \nodata & \nodata & \nodata & \nodata\\
$(V-K_2)$ & $-1.0$ & 1.366 & 4.474 & 86.0358 & -65.4928 & 10.8901 & \nodata & \nodata & \nodata & \nodata\\
$(V-K_2)$ & $-2.0$ & 1.334 & 3.549 & -6.96153 & 14.3298 & \nodata & \nodata & \nodata & \nodata & \nodata\\
$(V-K_2)$ & $-3.0$ & 1.258 & 2.768 & -943.925 & 1497.64 & -795.867 & 138.965 & \nodata & \nodata & \nodata\\ \hline
$(V_T-K_2)$ & $+0.0$ & 1.107 & 3.944 & -37.2128 & 31.2900 & -6.72743 & \nodata & \nodata & \nodata & \nodata\\
$(V_T-K_2)$ & $-1.0$ & 1.403 & 3.157 & -193.512 & 166.183 & -33.2781 & \nodata & \nodata & \nodata & \nodata\\
$(V_T-K_2)$ & $-2.0$ & 1.339 & 3.750 & -2.02136 & \nodata & \nodata & \nodata & \nodata & \nodata & \nodata\\
$(V_T-K_2)$ & $-3.0$ & 1.668 & 2.722 & 8.06982 & \nodata & \nodata & \nodata & \nodata & \nodata & \nodata 
\enddata
\tablenotetext{1}{Metallicity bins coded as: $\feh=+0.0$ ($-0.5<\feh<+0.5$), $\feh=-1.0$ ($-1.5<\feh\leq-0.5$), $\feh=-2.0$ ($-2.5<\feh\leq-1.5$), $\feh=-3.0$ ($-4.0<\feh\leq-2.5$).}
\label{t:pG}
\end{deluxetable*}

The number of stars in the dwarf calibrations is always larger than 120. The standard deviations range from 50~K (for $V-K_2$) to 121 K (for $Y-V$). For giants the number of stars in the calibrations amount from 90 to 270, while the standard deviations range from 28 K (for $V-K_2$) to 82~K (for $B_T-V_T$). The standard deviations are in general lower than those obtained by AAM, which is a consequence of adopting more accurate input parameters. For the $(B-V)$ calibration, for example, AAM obtained dispersions of 130~K for dwarfs and 100~K for giants. Our values are 88~K and 51~K, respectively. The general trends in the $\teff$ vs. color planes in common with AAM, however, are very similar (see~\S5). 

\begin{deluxetable*}{lccccccccccccc}
\tabletypesize{\scriptsize}
\tablecaption{Intrinsic $(B-V)$, $(b-y)$ and $(Y-V)$ Colors}
\tablehead{ & \multicolumn{4}{c}{$(B-V)$} & \multicolumn{4}{c}{$(b-y)$} & \multicolumn{4}{c}{$(Y-V)$} \\ \colhead{$\teff$\ / $\feh=$} & \colhead{$+0.0$} & \colhead{$-1.0$} & \colhead{$-2.0$} & \colhead{$-3.0$} & \colhead{$+0.0$} & \colhead{$-1.0$} & \colhead{$-2.0$} & \colhead{$-3.0$} & \colhead{$+0.0$} & \colhead{$-1.0$} & \colhead{$-2.0$} & \colhead{$-3.0$}}
\startdata 
\cutinhead{Dwarf Stars} 
4000	&	1.383	&	1.196	&	\nodata	&	\nodata	&	0.768	&	\nodata	&	\nodata	&	\nodata	&	0.925	&	\nodata	&	\nodata	&	\nodata	\\
4250	&	1.236	&	1.055	&	1.026	&	0.975	&	0.694	&	0.647	&	0.656	&	\nodata	&	0.861	&	\nodata	&	\nodata	&	\nodata	\\
4500	&	1.104	&	0.942	&	0.911	&	0.880	&	0.632	&	0.577	&	0.593	&	\nodata	&	0.785	&	\nodata	&	\nodata	&	\nodata	\\
4750	&	0.986	&	0.844	&	0.811	&	0.795	&	0.577	&	0.524	&	0.536	&	\nodata	&	0.724	&	\nodata	&	0.695	&	\nodata	\\
5000	&	0.882	&	0.757	&	0.721	&	0.718	&	0.526	&	0.479	&	0.489	&	\nodata	&	0.674	&	0.650	&	0.649	&	\nodata	\\
5250	&	0.788	&	0.678	&	0.642	&	0.649	&	0.480	&	0.439	&	0.445	&	\nodata	&	0.632	&	0.605	&	0.606	&	0.620	\\
5500	&	0.703	&	0.605	&	0.570	&	0.586	&	0.437	&	0.402	&	0.406	&	0.442	&	0.595	&	0.566	&	0.568	&	0.579	\\
5750	&	0.627	&	0.538	&	0.507	&	0.528	&	0.398	&	0.369	&	0.371	&	0.403	&	0.559	&	0.533	&	0.533	&	0.542	\\
6000	&	0.558	&	0.475	&	0.448	&	0.475	&	0.360	&	0.337	&	0.340	&	0.370	&	0.526	&	0.502	&	0.503	&	0.511	\\
6250	&	0.497	&	0.418	&	0.396	&	0.425	&	0.326	&	0.306	&	0.311	&	0.339	&	0.493	&	0.475	&	0.474	&	0.483	\\
6500	&	0.440	&	0.365	&	0.348	&	0.381	&	0.294	&	0.276	&	\nodata	&	0.311	&	0.461	&	\nodata	&	\nodata	&	0.458	\\
6750	&	0.388	&	0.316	&	\nodata	&	\nodata	&	0.264	&	0.246	&	\nodata	&	0.286	&	0.427	&	\nodata	&	\nodata	&	\nodata	\\
7000	&	0.341	&	\nodata	&	\nodata	&	\nodata	&	\nodata	&	\nodata	&	\nodata	&	\nodata	&	\nodata	&	\nodata	&	\nodata	&	\nodata	\\ 
\cutinhead{Giant Stars} 
3750	&	1.629	&	\nodata	&	\nodata	&	\nodata	&	1.016	&	\nodata	&	\nodata	&	\nodata	&	1.223	&	\nodata	&	\nodata	&	\nodata	\\
4000	&	1.481	&	1.405	&	\nodata	&	\nodata	&	0.906	&	\nodata	&	\nodata	&	\nodata	&	1.072	&	\nodata	&	\nodata	&	\nodata	\\
4250	&	1.324	&	1.218	&	1.198	&	\nodata	&	0.806	&	0.750	&	\nodata	&	\nodata	&	0.957	&	0.902	&	\nodata	&	\nodata	\\
4500	&	1.171	&	1.053	&	0.987	&	0.980	&	0.716	&	0.648	&	\nodata	&	0.674	&	0.863	&	0.815	&	\nodata	&	\nodata	\\
4750	&	1.032	&	0.908	&	0.830	&	0.780	&	0.634	&	0.571	&	0.569	&	0.590	&	0.782	&	0.737	&	0.712	&	0.742	\\
5000	&	0.909	&	0.778	&	0.701	&	\nodata	&	0.561	&	0.507	&	0.497	&	0.515	&	0.712	&	0.667	&	0.642	&	0.664	\\
5250	&	0.802	&	\nodata	&	\nodata	&	\nodata	&	0.497	&	0.453	&	0.442	&	0.446	&	0.649	&	0.603	&	0.589	&	0.605	\\
5500	&	0.707	&	\nodata	&	\nodata	&	\nodata	&	0.437	&	0.404	&	0.397	&	\nodata	&	0.594	&	\nodata	&	0.544	&	0.556	\\
5750	&	0.623	&	\nodata	&	\nodata	&	\nodata	&	0.385	&	0.360	&	\nodata	&	\nodata	&	0.544	&	\nodata	&	\nodata	&	0.513	\\
6000	&	0.547	&	\nodata	&	\nodata	&	\nodata	&	0.338	&	0.320	&	\nodata	&	\nodata	&	0.500	&	\nodata	&	\nodata	&	\nodata	\\
6250	&	0.478	&	\nodata	&	\nodata	&	\nodata	&	0.295	&	\nodata	&	\nodata	&	\nodata	&	0.459	&	\nodata	&	\nodata	&	\nodata	\\
6500	&	0.416	&	\nodata	&	\nodata	&	\nodata	&	0.255	&	\nodata	&	\nodata	&	\nodata	&	0.423	&	\nodata	&	\nodata	&	\nodata	\\
6750	&	0.358	&	\nodata	&	\nodata	&	\nodata	&	0.220	&	\nodata	&	\nodata	&	\nodata	&	0.389	&	\nodata	&	\nodata	&	\nodata	\\
7000	&	0.305	&	\nodata	&	\nodata	&	\nodata	&	0.187	&	\nodata	&	\nodata	&	\nodata	&	0.358	&	\nodata	&	\nodata	&	\nodata	\\
7250	&	0.256	&	\nodata	&	\nodata	&	\nodata	&	0.157	&	\nodata	&	\nodata	&	\nodata	&	0.330	&	\nodata	&	\nodata	&	\nodata	\\
7500	&	0.210	&	\nodata	&	\nodata	&	\nodata	&	0.130	&	\nodata	&	\nodata	&	\nodata	&	0.304	&	\nodata	&	\nodata	&	\nodata	\\
7750	&	0.166	&	\nodata	&	\nodata	&	\nodata	&	0.104	&	\nodata	&	\nodata	&	\nodata	&	0.280	&	\nodata	&	\nodata	&	\nodata	\\
8000	&	\nodata	&	\nodata	&	\nodata	&	\nodata	&	0.080	&	\nodata	&	\nodata	&	\nodata	&	0.258	&	\nodata	&	\nodata	&	\nodata	\\
8250	&	\nodata	&	\nodata	&	\nodata	&	\nodata	&	0.058	&	\nodata	&	\nodata	&	\nodata	&	0.237	&	\nodata	&	\nodata	&	\nodata	

\enddata
\label{t:esc1}
\end{deluxetable*}

\begin{deluxetable*}{lccccccccccccc}
\tabletypesize{\scriptsize}
\tablecaption{Intrinsic $(V-S)$, $(B_2-V_1)$ and $(B_2-G)$ Colors}
\tablehead{ & \multicolumn{4}{c}{$(V-S)$} & \multicolumn{4}{c}{$(B_2-V_1)$} & \multicolumn{4}{c}{$(B_2-G)$} \\ \colhead{$\teff$ \ / $\feh=$} & \colhead{$+0.0$} & \colhead{$-1.0$} & \colhead{$-2.0$} & \colhead{$-3.0$} & \colhead{$+0.0$} & \colhead{$-1.0$} & \colhead{$-2.0$} & \colhead{$-3.0$} & \colhead{$+0.0$} & \colhead{$-1.0$} & \colhead{$-2.0$} & \colhead{$-3.0$}}
\startdata
\cutinhead{Dwarf Stars}
4000	&	1.038	&	\nodata	&	\nodata	&	\nodata	&	\nodata	&	\nodata	&	\nodata	&	\nodata	&	1.005	&	\nodata	&	\nodata	&	\nodata	\\
4250	&	0.928	&	\nodata	&	\nodata	&	\nodata	&	0.806	&	\nodata	&	\nodata	&	\nodata	&	0.791	&	\nodata	&	0.530	&	\nodata	\\
4500	&	0.842	&	\nodata	&	0.755	&	\nodata	&	0.711	&	\nodata	&	\nodata	&	\nodata	&	0.626	&	0.460	&	0.418	&	\nodata	\\
4750	&	0.770	&	\nodata	&	0.698	&	\nodata	&	0.631	&	0.561	&	0.553	&	\nodata	&	0.487	&	0.350	&	0.316	&	\nodata	\\
5000	&	0.705	&	0.668	&	0.647	&	\nodata	&	0.559	&	0.494	&	0.484	&	\nodata	&	0.365	&	0.253	&	0.223	&	\nodata	\\
5250	&	0.648	&	0.614	&	0.602	&	\nodata	&	0.494	&	0.434	&	0.422	&	0.419	&	0.259	&	0.165	&	0.138	&	0.117	\\
5500	&	0.597	&	0.566	&	0.560	&	0.558	&	0.432	&	0.378	&	0.365	&	0.370	&	0.163	&	0.085	&	0.060	&	0.037	\\
5750	&	0.551	&	0.523	&	0.524	&	0.525	&	0.374	&	0.326	&	0.314	&	0.325	&	0.076	&	0.010	&	-0.010	&	-0.023	\\
6000	&	0.509	&	0.485	&	0.490	&	0.496	&	0.320	&	0.277	&	0.267	&	0.284	&	-0.002	&	-0.059	&	-0.074	&	-0.072	\\
6250	&	0.472	&	0.450	&	0.459	&	0.469	&	0.269	&	0.230	&	0.224	&	0.246	&	-0.073	&	-0.123	&	-0.133	&	-0.115	\\
6500	&	0.437	&	0.419	&	\nodata	&	0.444	&	0.221	&	0.187	&	0.186	&	0.212	&	-0.138	&	-0.182	&	-0.187	&	-0.152	\\
6750	&	0.407	&	\nodata	&	\nodata	&	\nodata	&	0.176	&	0.146	&	\nodata	&	\nodata	&	-0.197	&	-0.238	&	\nodata	&	\nodata	\\
7000	&	0.379	&	\nodata	&	\nodata	&	\nodata	&	0.134	&	\nodata	&	\nodata	&	\nodata	&	-0.251	&	\nodata	&	\nodata	&	\nodata	\\ 
\cutinhead{Giant Stars}
3750	&	1.147	&	\nodata	&	\nodata	&	\nodata	&	1.253	&	\nodata	&	\nodata	&	\nodata	&	\nodata	&	\nodata	&	\nodata	&	\nodata	\\
4000	&	0.999	&	0.953	&	\nodata	&	\nodata	&	1.081	&	1.017	&	\nodata	&	\nodata	&	1.038	&	0.944	&	\nodata	&	\nodata	\\
4250	&	0.892	&	0.862	&	0.888	&	\nodata	&	0.934	&	0.870	&	0.941	&	\nodata	&	0.835	&	0.739	&	0.777	&	\nodata	\\
4500	&	0.809	&	0.783	&	0.795	&	\nodata	&	0.807	&	0.741	&	0.731	&	\nodata	&	0.662	&	0.560	&	0.550	&	\nodata	\\
4750	&	0.739	&	0.714	&	0.716	&	0.728	&	0.696	&	0.627	&	0.590	&	0.554	&	0.511	&	0.404	&	0.358	&	0.318	\\
5000	&	0.679	&	0.652	&	0.647	&	0.658	&	0.599	&	0.526	&	0.478	&	0.457	&	0.378	&	0.265	&	0.192	&	0.181	\\
5250	&	0.625	&	0.598	&	0.587	&	0.596	&	0.512	&	0.436	&	0.384	&	\nodata	&	0.262	&	\nodata	&	\nodata	&	\nodata	\\
5500	&	0.577	&	0.548	&	0.535	&	\nodata	&	0.435	&	\nodata	&	\nodata	&	\nodata	&	0.156	&	\nodata	&	\nodata	&	\nodata	\\
5750	&	0.532	&	\nodata	&	\nodata	&	\nodata	&	0.365	&	\nodata	&	\nodata	&	\nodata	&	0.063	&	\nodata	&	\nodata	&	\nodata	\\
6000	&	0.491	&	\nodata	&	\nodata	&	\nodata	&	0.303	&	\nodata	&	\nodata	&	\nodata	&	-0.022	&	\nodata	&	\nodata	&	\nodata	\\
6250	&	0.453	&	\nodata	&	\nodata	&	\nodata	&	0.246	&	\nodata	&	\nodata	&	\nodata	&	-0.099	&	\nodata	&	\nodata	&	\nodata	\\
6500	&	0.418	&	\nodata	&	\nodata	&	\nodata	&	0.194	&	\nodata	&	\nodata	&	\nodata	&	-0.169	&	\nodata	&	\nodata	&	\nodata	\\
6750	&	0.384	&	\nodata	&	\nodata	&	\nodata	&	0.146	&	\nodata	&	\nodata	&	\nodata	&	-0.233	&	\nodata	&	\nodata	&	\nodata	\\
7000	&	0.353	&	\nodata	&	\nodata	&	\nodata	&	0.102	&	\nodata	&	\nodata	&	\nodata	&	-0.292	&	\nodata	&	\nodata	&	\nodata	\\
7250	&	0.324	&	\nodata	&	\nodata	&	\nodata	&	0.062	&	\nodata	&	\nodata	&	\nodata	&	-0.346	&	\nodata	&	\nodata	&	\nodata	\\
7500	&	0.297	&	\nodata	&	\nodata	&	\nodata	&	0.025	&	\nodata	&	\nodata	&	\nodata	&	-0.396	&	\nodata	&	\nodata	&	\nodata	\\
7750	&	0.271	&	\nodata	&	\nodata	&	\nodata	&	-0.010	&	\nodata	&	\nodata	&	\nodata	&	-0.443	&	\nodata	&	\nodata	&	\nodata	\\
8000	&	\nodata	&	\nodata	&	\nodata	&	\nodata	&	-0.042	&	\nodata	&	\nodata	&	\nodata	&	-0.486	&	\nodata	&	\nodata	&	\nodata	\\
8250	&	\nodata	&	\nodata	&	\nodata	&	\nodata	&	-0.073	&	\nodata	&	\nodata	&	\nodata	&	-0.526	&	\nodata	&	\nodata	&	\nodata	

\enddata
\label{t:esc2}
\end{deluxetable*}

\begin{deluxetable*}{lccccccccccccc}
\tabletypesize{\scriptsize}
\tablecaption{Intrinsic $t$, $(V-R_C)$ and $(V-I_C)$ Colors}
\tablehead{ & \multicolumn{4}{c}{$t$} & \multicolumn{4}{c}{$(V-R_C)$} & \multicolumn{4}{c}{$(V-I_C)$} \\ \colhead{$\teff$ \ / $\feh=$} & \colhead{$+0.0$} & \colhead{$-1.0$} & \colhead{$-2.0$} & \colhead{$-3.0$} & \colhead{$+0.0$} & \colhead{$-1.0$} & \colhead{$-2.0$} & \colhead{$-3.0$} & \colhead{$+0.0$} & \colhead{$-1.0$} & \colhead{$-2.0$} & \colhead{$-3.0$}}
\startdata
\cutinhead{Dwarf Stars}
3750	&	\nodata	&	\nodata	&	\nodata	&	\nodata	&	\nodata	&	\nodata	&	\nodata	&	\nodata	&	1.713	&	\nodata	&	\nodata	&	\nodata	\\
4000	&	\nodata	&	\nodata	&	\nodata	&	\nodata	&	0.822	&	\nodata	&	\nodata	&	\nodata	&	1.581	&	\nodata	&	\nodata	&	\nodata	\\
4250	&	\nodata	&	\nodata	&	\nodata	&	\nodata	&	0.756	&	\nodata	&	\nodata	&	\nodata	&	1.421	&	\nodata	&	\nodata	&	\nodata	\\
4500	&	\nodata	&	\nodata	&	\nodata	&	\nodata	&	0.674	&	\nodata	&	\nodata	&	\nodata	&	1.225	&	\nodata	&	\nodata	&	\nodata	\\
4750	&	0.403	&	0.366	&	\nodata	&	\nodata	&	0.579	&	0.509	&	0.488	&	\nodata	&	1.059	&	0.980	&	0.958	&	\nodata	\\
5000	&	0.338	&	0.308	&	0.314	&	\nodata	&	0.498	&	0.454	&	0.440	&	\nodata	&	0.937	&	0.882	&	0.876	&	\nodata	\\
5250	&	0.277	&	0.252	&	0.259	&	0.269	&	0.440	&	0.407	&	0.399	&	\nodata	&	0.840	&	0.800	&	0.801	&	\nodata	\\
5500	&	0.220	&	0.200	&	0.206	&	0.220	&	0.393	&	0.366	&	0.362	&	\nodata	&	0.760	&	0.731	&	0.735	&	\nodata	\\
5750	&	0.165	&	0.149	&	0.157	&	0.174	&	0.355	&	0.330	&	0.329	&	\nodata	&	0.690	&	0.670	&	0.674	&	\nodata	\\
6000	&	0.112	&	0.100	&	0.109	&	0.131	&	0.321	&	0.298	&	0.300	&	0.318	&	0.626	&	0.616	&	0.619	&	\nodata	\\
6250	&	0.059	&	0.053	&	0.064	&	0.090	&	0.290	&	\nodata	&	0.273	&	0.291	&	0.569	&	\nodata	&	0.571	&	\nodata	\\
6500	&	0.007	&	0.007	&	0.020	&	0.051	&	0.261	&	\nodata	&	\nodata	&	0.267	&	0.515	&	\nodata	&	\nodata	&	\nodata	\\
6750	&	-0.046	&	-0.038	&	\nodata	&	\nodata	&	0.233	&	\nodata	&	\nodata	&	0.245	&	\nodata	&	\nodata	&	\nodata	&	\nodata	\\
7000	&	-0.100	&	\nodata	&	\nodata	&	\nodata	&	0.206	&	\nodata	&	\nodata	&	\nodata	&	\nodata	&	\nodata	&	\nodata	&	\nodata	\\ 
\cutinhead{Giant Stars}
3750	&	0.926	&	\nodata	&	\nodata	&	\nodata	&	0.948	&	\nodata	&	\nodata	&	\nodata	&	1.861	&	\nodata	&	\nodata	&	\nodata	\\
4000	&	0.781	&	\nodata	&	\nodata	&	\nodata	&	0.782	&	\nodata	&	\nodata	&	\nodata	&	1.492	&	\nodata	&	\nodata	&	\nodata	\\
4250	&	0.647	&	0.652	&	\nodata	&	\nodata	&	0.673	&	0.657	&	\nodata	&	\nodata	&	1.278	&	1.281	&	\nodata	&	\nodata	\\
4500	&	0.530	&	0.528	&	0.561	&	\nodata	&	0.592	&	0.568	&	0.577	&	\nodata	&	1.122	&	1.110	&	1.135	&	\nodata	\\
4750	&	0.430	&	0.424	&	0.440	&	0.439	&	0.527	&	0.498	&	0.497	&	0.496	&	0.998	&	0.977	&	0.987	&	1.006	\\
5000	&	0.346	&	0.335	&	0.337	&	0.332	&	0.472	&	0.440	&	0.432	&	0.424	&	0.897	&	0.870	&	\nodata	&	0.877	\\
5250	&	0.275	&	0.259	&	0.250	&	0.241	&	0.426	&	0.391	&	\nodata	&	\nodata	&	0.811	&	\nodata	&	\nodata	&	\nodata	\\
5500	&	0.213	&	0.193	&	0.175	&	\nodata	&	0.387	&	\nodata	&	\nodata	&	\nodata	&	0.738	&	\nodata	&	\nodata	&	\nodata	\\
5750	&	0.158	&	0.134	&	\nodata	&	\nodata	&	0.351	&	\nodata	&	\nodata	&	\nodata	&	0.673	&	\nodata	&	\nodata	&	\nodata	\\
6000	&	0.110	&	0.082	&	\nodata	&	\nodata	&	0.321	&	\nodata	&	\nodata	&	\nodata	&	0.617	&	\nodata	&	\nodata	&	\nodata	

\enddata
\label{t:esc3}
\end{deluxetable*}

\begin{deluxetable*}{lccccccccccccc}
\tabletypesize{\scriptsize}
\tablecaption{Intrinsic $(R_C-I_C)$, $C(42-45)$ and $C(42-48)$ Colors}
\tablehead{ & \multicolumn{4}{c}{$(R_C-I_C)$} & \multicolumn{4}{c}{$C(42-45)$} & \multicolumn{4}{c}{$C(42-48)$} \\ \colhead{$\teff$ \ / $\feh=$} & \colhead{$+0.0$} & \colhead{$-1.0$} & \colhead{$-2.0$} & \colhead{$-3.0$} & \colhead{$+0.0$} & \colhead{$-1.0$} & \colhead{$-2.0$} & \colhead{$-3.0$} & \colhead{$+0.0$} & \colhead{$-1.0$} & \colhead{$-2.0$} & \colhead{$-3.0$}}
\startdata
\cutinhead{Dwarf Stars}
3750	&	0.826	&	\nodata	&	\nodata	&	\nodata	&	\nodata	&	\nodata	&	\nodata	&	\nodata	&	\nodata	&	\nodata	&	\nodata	&	\nodata	\\
4000	&	0.745	&	0.693	&	\nodata	&	\nodata	&	1.418	&	\nodata	&	\nodata	&	\nodata	&	2.620	&	\nodata	&	\nodata	&	\nodata	\\
4250	&	0.632	&	0.600	&	\nodata	&	\nodata	&	1.313	&	\nodata	&	\nodata	&	\nodata	&	2.461	&	\nodata	&	\nodata	&	\nodata	\\
4500	&	0.549	&	0.532	&	0.505	&	\nodata	&	1.187	&	\nodata	&	\nodata	&	\nodata	&	2.298	&	\nodata	&	\nodata	&	\nodata	\\
4750	&	0.489	&	0.479	&	0.468	&	\nodata	&	1.047	&	\nodata	&	\nodata	&	\nodata	&	2.141	&	\nodata	&	\nodata	&	\nodata	\\
5000	&	0.442	&	0.435	&	0.436	&	\nodata	&	0.915	&	0.761	&	\nodata	&	\nodata	&	1.997	&	1.843	&	\nodata	&	\nodata	\\
5250	&	0.402	&	0.397	&	0.406	&	\nodata	&	0.804	&	0.681	&	\nodata	&	\nodata	&	1.871	&	1.738	&	\nodata	&	\nodata	\\
5500	&	0.366	&	0.366	&	0.378	&	\nodata	&	0.714	&	0.611	&	\nodata	&	\nodata	&	1.759	&	1.644	&	\nodata	&	\nodata	\\
5750	&	0.333	&	0.338	&	0.350	&	0.362	&	0.641	&	0.550	&	\nodata	&	\nodata	&	1.660	&	1.558	&	1.504	&	\nodata	\\
6000	&	0.305	&	0.314	&	0.322	&	0.339	&	0.577	&	0.495	&	\nodata	&	\nodata	&	1.571	&	1.480	&	1.442	&	\nodata	\\
6250	&	0.277	&	\nodata	&	0.295	&	0.317	&	0.523	&	\nodata	&	\nodata	&	\nodata	&	1.491	&	\nodata	&	\nodata	&	\nodata	\\
6500	&	0.252	&	\nodata	&	\nodata	&	0.299	&	0.475	&	\nodata	&	\nodata	&	\nodata	&	1.418	&	\nodata	&	\nodata	&	\nodata	\\
6750	&	\nodata	&	\nodata	&	\nodata	&	\nodata	&	\nodata	&	\nodata	&	\nodata	&	\nodata	&	1.352	&	\nodata	&	\nodata	&	\nodata	\\
7000	&	\nodata	&	\nodata	&	\nodata	&	\nodata	&	\nodata	&	\nodata	&	\nodata	&	\nodata	&	1.291	&	\nodata	&	\nodata	&	\nodata	\\
\cutinhead{Giant Stars}
3750	&	\nodata	&	\nodata	&	\nodata	&	\nodata	&	\nodata	&	1.253	&	\nodata	&	\nodata	&	\nodata	&	\nodata	&	\nodata	&	\nodata	\\
4000	&	0.720	&	0.711	&	\nodata	&	\nodata	&	1.322	&	1.093	&	\nodata	&	\nodata	&	2.676	&	2.441	&	\nodata	&	\nodata	\\
4250	&	0.618	&	0.614	&	0.632	&	\nodata	&	1.149	&	0.962	&	0.808	&	\nodata	&	2.456	&	2.229	&	2.093	&	\nodata	\\
4500	&	0.541	&	0.540	&	0.560	&	\nodata	&	1.009	&	0.853	&	0.715	&	0.624	&	2.252	&	2.052	&	1.917	&	\nodata	\\
4750	&	0.479	&	0.480	&	0.501	&	0.504	&	0.892	&	0.758	&	0.634	&	0.534	&	2.080	&	1.903	&	1.769	&	1.688	\\
5000	&	0.427	&	0.431	&	0.451	&	0.458	&	0.793	&	0.677	&	0.563	&	\nodata	&	1.938	&	1.774	&	1.643	&	1.583	\\
5250	&	\nodata	&	0.388	&	\nodata	&	0.418	&	0.708	&	0.605	&	0.502	&	\nodata	&	1.817	&	1.662	&	1.533	&	\nodata	\\
5500	&	\nodata	&	\nodata	&	\nodata	&	0.384	&	0.633	&	0.542	&	0.446	&	\nodata	&	1.715	&	1.562	&	\nodata	&	\nodata	\\
5750	&	\nodata	&	\nodata	&	\nodata	&	\nodata	&	0.567	&	0.486	&	\nodata	&	\nodata	&	1.627	&	1.474	&	\nodata	&	\nodata	\\
6000	&	\nodata	&	\nodata	&	\nodata	&	\nodata	&	0.508	&	0.435	&	\nodata	&	\nodata	&	1.547	&	\nodata	&	\nodata	&	\nodata	\\
6250	&	\nodata	&	\nodata	&	\nodata	&	\nodata	&	0.455	&	\nodata	&	\nodata	&	\nodata	&	\nodata	&	\nodata	&	\nodata	&	\nodata	

\enddata
\label{t:esc4}
\end{deluxetable*}

\begin{deluxetable*}{lccccccccccccc}
\tabletypesize{\scriptsize}
\tablecaption{Intrinsic $(B_T-V_T)$, $(V-J_2)$ and $(V-H_2)$ Colors}
\tablehead{ & \multicolumn{4}{c}{$(B_T-V_T)$} & \multicolumn{4}{c}{$(V-J_2)$} & \multicolumn{4}{c}{$(V-H_2)$} \\ \colhead{$\teff$ \ / $\feh=$} & \colhead{$+0.0$} & \colhead{$-1.0$} & \colhead{$-2.0$} & \colhead{$-3.0$} & \colhead{$+0.0$} & \colhead{$-1.0$} & \colhead{$-2.0$} & \colhead{$-3.0$} & \colhead{$+0.0$} & \colhead{$-1.0$} & \colhead{$-2.0$} & \colhead{$-3.0$}}
\startdata 
\cutinhead{Dwarf Stars}
4000	&	1.563	&	1.464	&	\nodata	&	\nodata	&	\nodata	&	\nodata	&	\nodata	&	\nodata	&	\nodata	&	\nodata	&	\nodata	&	\nodata	\\
4250	&	1.427	&	1.297	&	\nodata	&	\nodata	&	2.322	&	2.072	&	1.950	&	1.886	&	2.899	&	\nodata	&	2.469	&	2.342	\\
4500	&	1.295	&	1.145	&	\nodata	&	\nodata	&	2.040	&	1.875	&	1.795	&	1.758	&	2.562	&	2.413	&	2.290	&	2.172	\\
4750	&	1.165	&	1.009	&	0.912	&	\nodata	&	1.793	&	1.695	&	1.647	&	1.633	&	2.267	&	2.163	&	2.104	&	2.013	\\
5000	&	1.035	&	0.886	&	0.825	&	\nodata	&	1.587	&	1.534	&	1.507	&	1.512	&	2.010	&	1.947	&	1.922	&	1.863	\\
5250	&	0.910	&	0.775	&	0.704	&	\nodata	&	1.417	&	1.390	&	1.377	&	1.398	&	1.784	&	1.760	&	1.751	&	1.725	\\
5500	&	0.797	&	0.674	&	0.622	&	\nodata	&	1.275	&	1.262	&	1.259	&	1.291	&	1.587	&	1.594	&	1.595	&	1.597	\\
5750	&	0.698	&	0.582	&	0.554	&	\nodata	&	1.154	&	1.147	&	1.153	&	1.192	&	1.414	&	1.448	&	1.459	&	1.480	\\
6000	&	0.613	&	0.498	&	0.486	&	0.474	&	1.048	&	1.045	&	1.057	&	1.102	&	1.260	&	1.317	&	1.337	&	1.372	\\
6250	&	0.541	&	0.420	&	\nodata	&	0.426	&	0.956	&	0.955	&	0.970	&	1.020	&	1.121	&	1.199	&	1.229	&	1.275	\\
6500	&	0.478	&	\nodata	&	\nodata	&	\nodata	&	0.875	&	0.873	&	\nodata	&	0.945	&	0.998	&	1.092	&	1.134	&	1.184	\\
6750	&	0.422	&	\nodata	&	\nodata	&	\nodata	&	\nodata	&	\nodata	&	\nodata	&	\nodata	&	0.886	&	\nodata	&	\nodata	&	1.101	\\
7000	&	0.373	&	\nodata	&	\nodata	&	\nodata	&	\nodata	&	\nodata	&	\nodata	&	\nodata	&	\nodata	&	\nodata	&	\nodata	&	\nodata	\\ 
\cutinhead{Giant Stars}
3750	&	1.934	&	\nodata	&	\nodata	&	\nodata	&	\nodata	&	2.886	&	\nodata	&	\nodata	&	\nodata	&	3.706	&	\nodata	&	\nodata	\\
4000	&	1.773	&	1.630	&	\nodata	&	\nodata	&	\nodata	&	2.428	&	2.607	&	\nodata	&	\nodata	&	3.173	&	3.257	&	\nodata	\\
4250	&	1.592	&	1.386	&	\nodata	&	\nodata	&	2.196	&	2.114	&	2.218	&	\nodata	&	2.832	&	2.775	&	2.830	&	\nodata	\\
4500	&	1.400	&	1.190	&	1.106	&	\nodata	&	1.939	&	1.874	&	1.938	&	2.001	&	2.514	&	2.456	&	2.493	&	2.494	\\
4750	&	1.213	&	1.022	&	0.906	&	0.795	&	1.736	&	1.679	&	1.719	&	1.763	&	2.239	&	2.191	&	2.215	&	2.221	\\
5000	&	1.045	&	0.878	&	0.761	&	0.641	&	1.571	&	1.516	&	1.541	&	1.572	&	1.997	&	1.966	&	1.981	&	1.990	\\
5250	&	0.904	&	0.750	&	0.645	&	0.525	&	1.431	&	1.379	&	1.392	&	1.412	&	1.785	&	1.772	&	1.780	&	1.791	\\
5500	&	0.784	&	0.637	&	0.543	&	\nodata	&	1.311	&	1.260	&	1.263	&	1.278	&	1.600	&	1.602	&	1.606	&	1.618	\\
5750	&	0.682	&	0.534	&	\nodata	&	\nodata	&	\nodata	&	1.155	&	1.151	&	1.160	&	1.436	&	1.451	&	1.451	&	1.466	\\
6000	&	0.595	&	0.441	&	\nodata	&	\nodata	&	\nodata	&	1.063	&	1.052	&	1.058	&	1.292	&	1.317	&	1.315	&	1.330	\\
6250	&	0.518	&	\nodata	&	\nodata	&	\nodata	&	\nodata	&	\nodata	&	\nodata	&	\nodata	&	\nodata	&	\nodata	&	\nodata	&	\nodata	\\
6500	&	0.450	&	\nodata	&	\nodata	&	\nodata	&	\nodata	&	\nodata	&	\nodata	&	\nodata	&	\nodata	&	\nodata	&	\nodata	&	\nodata	\\
6750	&	0.389	&	\nodata	&	\nodata	&	\nodata	&	\nodata	&	\nodata	&	\nodata	&	\nodata	&	\nodata	&	\nodata	&	\nodata	&	\nodata	\\
7000	&	0.334	&	\nodata	&	\nodata	&	\nodata	&	\nodata	&	\nodata	&	\nodata	&	\nodata	&	\nodata	&	\nodata	&	\nodata	&	\nodata	\\
7250	&	0.284	&	\nodata	&	\nodata	&	\nodata	&	\nodata	&	\nodata	&	\nodata	&	\nodata	&	\nodata	&	\nodata	&	\nodata	&	\nodata	\\
7500	&	0.237	&	\nodata	&	\nodata	&	\nodata	&	\nodata	&	\nodata	&	\nodata	&	\nodata	&	\nodata	&	\nodata	&	\nodata	&	\nodata	\\
7750	&	0.194	&	\nodata	&	\nodata	&	\nodata	&	\nodata	&	\nodata	&	\nodata	&	\nodata	&	\nodata	&	\nodata	&	\nodata	&	\nodata	\\
8000	&	0.154	&	\nodata	&	\nodata	&	\nodata	&	\nodata	&	\nodata	&	\nodata	&	\nodata	&	\nodata	&	\nodata	&	\nodata	&	\nodata	

\enddata
\label{t:esc5}
\end{deluxetable*}

\begin{deluxetable*}{lccccccccc}
\tabletypesize{\scriptsize}
\tablecaption{Intrinsic $(V-K_2)$ and $(V_T-K_2)$ Colors}
\tablehead{ & \multicolumn{4}{c}{$(V-K_2)$} & \multicolumn{4}{c}{$(V_T-K_2)$} \\ \colhead{$\teff$ \ / $\feh=$} & \colhead{$+0.0$} & \colhead{$-1.0$} & \colhead{$-2.0$} & \colhead{$-3.0$} & \colhead{$+0.0$} & \colhead{$-1.0$} & \colhead{$-2.0$} & \colhead{$-3.0$} }
\startdata 
\cutinhead{Dwarf Stars}
4250	&	3.106	&	\nodata	&	2.609	&	2.516	&	3.284	&	\nodata	&	\nodata	&	\nodata	\\
4500	&	2.728	&	2.538	&	2.437	&	2.352	&	2.850	&	\nodata	&	\nodata	&	\nodata	\\
4750	&	2.374	&	2.275	&	2.234	&	2.182	&	2.476	&	2.441	&	2.402	&	\nodata	\\
5000	&	2.097	&	2.046	&	2.022	&	2.012	&	2.189	&	2.162	&	2.108	&	\nodata	\\
5250	&	1.874	&	1.845	&	1.829	&	1.847	&	1.957	&	1.932	&	1.906	&	\nodata	\\
5500	&	1.682	&	1.669	&	1.662	&	1.693	&	1.758	&	1.738	&	1.738	&	\nodata	\\
5750	&	1.512	&	1.511	&	1.515	&	1.553	&	1.580	&	1.568	&	1.587	&	1.620	\\
6000	&	1.357	&	1.370	&	1.386	&	1.426	&	1.415	&	1.420	&	1.447	&	1.470	\\
6250	&	1.212	&	1.243	&	1.268	&	1.312	&	1.261	&	1.288	&	1.318	&	1.340	\\
6500	&	1.074	&	1.126	&	1.161	&	1.210	&	1.115	&	1.169	&	\nodata	&	1.227	\\
6750	&	0.942	&	\nodata	&	\nodata	&	\nodata	&	0.974	&	\nodata	&	\nodata	&	\nodata	\\ 
\cutinhead{Giant Stars}
3750	&	\nodata	&	3.911	&	\nodata	&	\nodata	&	\nodata	&	\nodata	&	\nodata	&	\nodata	\\
4000	&	\nodata	&	3.322	&	3.436	&	\nodata	&	3.626	&	\nodata	&	3.565	&	\nodata	\\
4250	&	3.005	&	2.893	&	2.978	&	\nodata	&	3.148	&	3.122	&	3.106	&	\nodata	\\
4500	&	2.625	&	2.554	&	2.617	&	2.660	&	2.759	&	2.754	&	2.734	&	\nodata	\\
4750	&	2.318	&	2.274	&	2.323	&	2.328	&	2.436	&	2.440	&	2.427	&	2.462	\\
5000	&	2.064	&	2.039	&	2.076	&	2.085	&	2.164	&	2.168	&	2.166	&	2.205	\\
5250	&	1.847	&	1.834	&	1.863	&	1.883	&	1.928	&	1.930	&	1.940	&	1.981	\\
5500	&	1.660	&	1.657	&	1.679	&	1.706	&	1.723	&	1.723	&	1.744	&	1.786	\\
5750	&	1.497	&	1.500	&	1.518	&	1.547	&	1.542	&	1.539	&	1.570	&	\nodata	\\
6000	&	1.351	&	\nodata	&	1.374	&	1.401	&	1.380	&	\nodata	&	1.415	&	\nodata	\\
6250	&	\nodata	&	\nodata	&	\nodata	&	1.268	&	1.236	&	\nodata	&	\nodata	&	\nodata	

\enddata
\label{t:esc6}
\end{deluxetable*}

\bigskip

\section{The empirical IRFM temperature scale} \label{sect:scale}

\subsection{Intrinsic colors of dwarfs and giants}

By numerically inverting Eq.~(\ref{eq:teff}) for a given $\teff,\feh$ pair, grids of colors in the $(\teff,\feh)$ space were derived. This procedure is safer and easier than calibrating the relations from the data, since it guarantees a single correspondence between the three quantities involved, i.e. the grids and the calibration formulae produce the same results. The intrinsic colors of both dwarfs and giants as a function of $\teff$ and $\feh$ are given in Tables \ref{t:esc1}-\ref{t:esc6}. 

\subsection{The IRFM vs. direct temperature scales}

In Part I we compared the IRFM temperatures with the direct temperatures for both dwarf and giant stars. The comparison showed that the zero points of both temperature scales are essentially equal. As a practical example of the use of the present calibrations, and to test their reliability, here we derive the temperatures from the colors of the same stars used for the comparison of direct and IRFM temperatures in Part~I. 

In Table~\ref{t:teffdir} we give the temperatures derived from angular diameter and bolometric flux measurements ($\teff^\mathrm{dir}$), as given in Part I for dwarfs (see Sect.~5.1 in Part~I for the references) and by Mozurkewich et~al. (2003) for giants, along with the temperatures we obtained with the IRFM in Part~I ($\teff^\mathrm{IRFM}$). The average of the temperatures obtained from the color calibrations are given as $\teff^\mathrm{cal}$. The number of colors used ($N$) for each star is also provided. More than eight color calibrations were used.

\begin{deluxetable*}{rrlllrl}
\tabletypesize{\scriptsize}
\tablecaption{Comparison of Direct and IRFM Temperatures\tablenotemark{1}}
\tablehead{\colhead{HD} & \colhead{$\feh$} & \colhead{$\teff^\mathrm{dir}$} & \colhead{$\teff^\mathrm{IRFM}$} & \colhead{$\teff^\mathrm{cal}$} & \colhead{$N$} & $\teff$ (adopted)}
\startdata
\cutinhead{Dwarf and Subgiant Stars}
10700	&	-0.54	&	5319	$\pm$	43	&	5372	$\pm$	65	&	5270	$\pm$	82	&	10	&	5304	$\pm$	105	\\
16160	&	-0.03	&	\nodata	&	4714	$\pm$	67	&	4781	$\pm$	42	&	12	&	4759	$\pm$	79	\\
22049	&	-0.12	&	5078	$\pm$	40	&	5015	$\pm$	56	&	4981	$\pm$	38	&	11	&	4992	$\pm$	68	\\
26965	&	-0.28	&	\nodata	&	5068	$\pm$	63	&	5099	$\pm$	25	&	10	&	5089	$\pm$	68	\\
61421	&	-0.01	&	6532	$\pm$	39	&	6591	$\pm$	73	&	6586	$\pm$	46	&	9	&	6588	$\pm$	86	\\
88230	&	-0.12	&	3967	$\pm$	63	&	3950	$\pm$	161	&	3985	$\pm$	73	&	9	&	3973	$\pm$	177	\\
121370	&	0.25	&	6081	$\pm$	47	&	6038	$\pm$	75	&	6034	$\pm$	47	&	8	&	6035	$\pm$	88	\\
128620	&	0.20	&	5771	$\pm$	23	&	5759	$\pm$	70	&	5736	$\pm$	19	&	4	&	5744	$\pm$	72	\\
128621	&	0.20	&	5178	$\pm$	23	&	5201	$\pm$	65	&	5103	$\pm$	20	&	6	&	5136	$\pm$	68	\\
131977	&	0.09	&	4469	$\pm$	57	&	4571	$\pm$	52	&	4545	$\pm$	26	&	5	&	4554	$\pm$	58	\\
198149	&	-0.18	&	4939	$\pm$	41	&	4907	$\pm$	54	&	4903	$\pm$	94	&	10	&	4904	$\pm$	108	\\
209100	&	-0.02	&	4527	$\pm$	29	&	4642	$\pm$	54	&	4605	$\pm$	20	&	11	&	4617	$\pm$	58	\\
209458	&	-0.01	&	\nodata	&	5993	$\pm$	71	&	5983	$\pm$	37	&	6	&	5986	$\pm$	80	\\ 
\cutinhead{Giant Stars}
3546	&	-0.72	&	\nodata			&	4935	$\pm$	52	&	4880	$\pm$	44	&	10	&	4898	$\pm$	68	\\
3627	&	0.17	&	4392	$\pm$	27	&	4343	$\pm$	45	&	4350	$\pm$	20	&	8	&	4348	$\pm$	49	\\
3712	&	-0.10	&	4602	$\pm$	29	&	4553	$\pm$	48	&	4516	$\pm$	51	&	7	&	4528	$\pm$	70	\\
6860	&	-0.07	&	\nodata	&	3824	$\pm$	41	&	3845	$\pm$	45	&	9	&	3838	$\pm$	61	\\
9927	&	-0.01	&	\nodata			&	4380	$\pm$	48	&	4298	$\pm$	60	&	8	&	4325	$\pm$	77	\\
10380	&	-0.27	&	\nodata			&	4132	$\pm$	46	&	4146	$\pm$	19	&	13	&	4141	$\pm$	50	\\
12533	&	-0.07	&	4254	$\pm$	27	&	4259	$\pm$	45	&	4195	$\pm$	50	&	4	&	4216	$\pm$	67	\\
12929	&	-0.25	&	4493	$\pm$	28	&	4501	$\pm$	50	&	4500	$\pm$	25	&	10	&	4500	$\pm$	56	\\
18884	&	0.00	&	\nodata	&	3718	$\pm$	46	&	3739	$\pm$	31	&	7	&	3732	$\pm$	55	\\
29139	&	-0.18	&	3871	$\pm$	24	&	3883	$\pm$	44	&	3901	$\pm$	33	&	10	&	3895	$\pm$	55	\\
62509	&	-0.02	&	4858	$\pm$	30	&	4833	$\pm$	50	&	4822	$\pm$	20	&	8	&	4826	$\pm$	54	\\
62721	&	-0.27	&	\nodata			&	3988	$\pm$	48	&	4005	$\pm$	32	&	5	&	3999	$\pm$	58	\\
76294	&	-0.01	&	\nodata			&	4817	$\pm$	50	&	4805	$\pm$	83	&	11	&	4809	$\pm$	97	\\
80493	&	-0.26	&	3836	$\pm$	24	&	3851	$\pm$	42	&	3868	$\pm$	14	&	7	&	3862	$\pm$	44	\\
94264	&	-0.20	&	\nodata			&	4670	$\pm$	51	&	4650	$\pm$	56	&	10	&	4657	$\pm$	76	\\
99998	&	-0.39	&	\nodata			&	3919	$\pm$	45	&	3896	$\pm$	10	&	8	&	3904	$\pm$	46	\\
102224	&	-0.44	&	\nodata			&	4378	$\pm$	46	&	4396	$\pm$	26	&	8	&	4390	$\pm$	53	\\
113226	&	0.11	&	4981	$\pm$	31	&	5049	$\pm$	59	&	4984	$\pm$	50	&	10	&	5006	$\pm$	77	\\
124897	&	-0.55	&	4226	$\pm$	29	&	4231	$\pm$	49	&	4283	$\pm$	42	&	13	&	4266	$\pm$	65	\\
135722	&	-0.40	&	4851	$\pm$	32	&	4834	$\pm$	50	&	4793	$\pm$	27	&	10	&	4807	$\pm$	57	\\
150997	&	-0.28	&	4841	$\pm$	36	&	4948	$\pm$	54	&	4922	$\pm$	34	&	9	&	4931	$\pm$	64	\\
164058	&	-0.15	&	4013	$\pm$	30	&	3927	$\pm$	42	&	3920	$\pm$	21	&	10	&	3922	$\pm$	47	\\
169414	&	-0.16	&	\nodata			&	4450	$\pm$	50	&	4462	$\pm$	26	&	7	&	4458	$\pm$	56	\\
181276	&	0.02	&	\nodata			&	4935	$\pm$	54	&	4942	$\pm$	21	&	10	&	4940	$\pm$	58	\\
189319	&	0.00	&	3858	$\pm$	24	&	3877	$\pm$	41	&	3859	$\pm$	15	&	6	&	3865	$\pm$	44	\\
197989	&	-0.12	&	4757	$\pm$	30	&	4710	$\pm$	52	&	4716	$\pm$	20	&	10	&	4714	$\pm$	56	\\
210745	&	0.25	&	4351	$\pm$	27	&	4482	$\pm$	51	&	4222	$\pm$	144	&	9	&	4309	$\pm$	153	\\
214868	&	-0.25	&	\nodata			&	4303	$\pm$	47	&	4265	$\pm$	25	&	7	&	4278	$\pm$	53	\\
217906	&	-0.11	&	\nodata	&	3648	$\pm$	43	&	3741	$\pm$	132	&	8	&	3710	$\pm$	139	\\
221115	&	0.04	&	\nodata			&	4980	$\pm$	63	&	4955	$\pm$	20	&	9	&	4963	$\pm$	66	\\
222107	&	-0.50	&	\nodata			&	4605	$\pm$	49	&	4650	$\pm$	34	&	10	&	4635	$\pm$	60 

\enddata
\tablenotetext{1}{Metallicities, direct, and IRFM temperatures as given in Part I. For dwarfs only the direct temperatures obtained with reliable angular diameter measurements are given; for giants the direct temperatures ($\teff>3800$~K only) are from Mozurkewich et~al. (2003). The temperatures obtained from $N$ color calibrations are also given. The last column corresponds to the suggested $\teff$ to adopt: $\teff=(2\teff^\mathrm{cal}+\teff^\mathrm{IRFM})/3$.}
\label{t:teffdir}
\end{deluxetable*}

As we showed in Part I, the photometric errors are important in our IRFM implementation. The IRFM temperatures we obtained in Part I depend on the quality of the $V$ and infrared magnitudes, which, in general (but not always), are reasonably accurate. With the color calibrations, the impact of photometric errors is removed as an average $\teff$ vs. color relation is constructed. When using a given color calibration, the systematic error in the obtained $\teff$ is mainly due to the error in that particular color, and only to a lesser extent due to the errors in the IRFM temperatures. If more than one color is used, the photometric errors from different colors may be reduced by taking an average $\teff$. Consequently, if an IRFM temperature is available for a given star, and a mean $\teff$ is obtained from its colors, the photometric temperature, provided that the number of colors used is large enough, will be more reliable. We have already shown this in the case of Arcturus (see Sect.~3.5 in RM04a).

Thus, whenever an IRFM temperature is available, we suggest the following temperature be adopted:
\begin{equation}
\teff=\frac{m\teff^\mathrm{cal}+n\teff^\mathrm{IRFM}}{m+n}\ ,\label{eq:teffadopt}
\end{equation}
where $m\ge n$. The exact values of $m$ and $n$ should be chosen depending on the quality of the colors and the IRFM temperature. The values we have used for the $\teff$~(adopted) in Table~\ref{t:teffdir} are $m=2$, $n=1$.

\begin{figure}
\epsscale{1.1}
\plotone{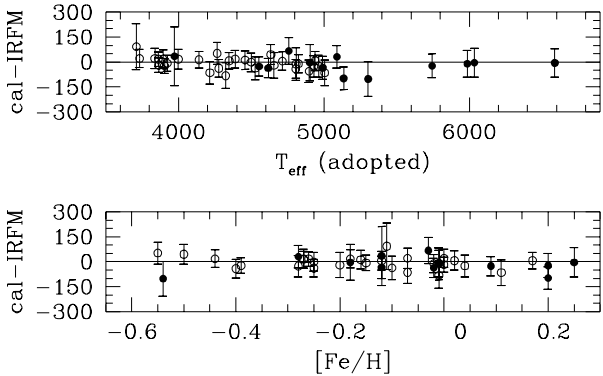} \caption{Difference between the temperatures from the color calibrations and the IRFM temperatures as a function of the adopted temperatures and metallicities of dwarfs (filled circles) and giants (open circles).} \label{fig:teffs1}
\end{figure}

The mean difference $\teff^\mathrm{cal}-\teff^\mathrm{IRFM}$ is $-16\pm48$~K for dwarfs and $-6\pm37$~K for giants. 
Fig. \ref{fig:teffs1} shows that no systematic errors are introduced by the calibrations, neither with $\teff$ nor $\feh$.

\begin{figure}
\epsscale{1.1}
\plotone{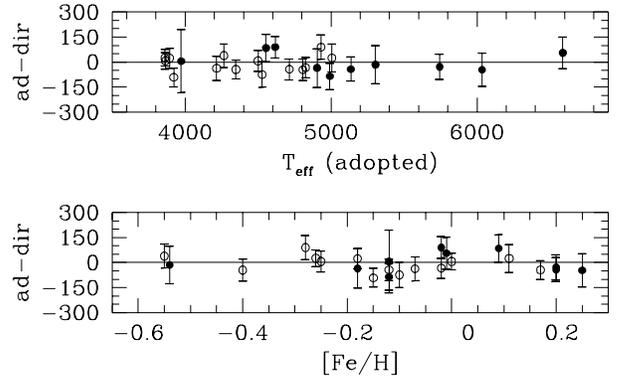} \caption{Difference between the adopted temperatures and direct temperatures as a function of the adopted temperatures and metallicities of dwarfs (filled circles) and giants (open circles).} \label{fig:teffs2}
\end{figure}

Given the reliability of the photometric temperatures, they may be safely combined with the IRFM temperatures according to Eq.~(\ref{eq:teffadopt}). When comparing the adopted temperatures (those obtained by combining the photometric and IRFM temperatures) with the direct ones, a mean difference of $-1\pm60$~K for dwarfs and $-11\pm50$~K for giants is obtained. As shown in Fig.~\ref{fig:teffs2}, no trends are observed with either $\teff$ or $\feh$. The dispersion in the mean differences for dwarfs is the same as the one found in Part I when only the IRFM temperatures were compared with the direct temperatures. However, when we consider the adopted temperatures, the zero point of the temperature scale is only 1~K below the direct one (it was 30~K when only the IRFM temperatures were used in Part~I). For giants, the zero points are still in excellent agreement (at the 10~K level), and the dispersion in the mean differences has been reduced by 10~K.

\subsection{The colors of the Sun}

Interpolation from Tables~\ref{t:esc1}-\ref{t:esc6} at $\teff=5777$ K and $\feh=0.0$ allowed us to derive colors representative of a solar-twin star. They are given in Table~\ref{t:sun} along with the colors of five solar analogs from the list of Soubiran \& Triaud (2004). The metallicities given by Soubiran \& Triaud have been adopted to derive the photometric temperatures of these stars, which are also given in Table~\ref{t:sun}.

\begin{deluxetable*}{rrrrrrr}
\tablecaption{Colors of the Sun and Solar Analogs}
\tablehead{\colhead{color} & \colhead{Sun} & \colhead{HD 146233} & \colhead{HD 10307} & \colhead{HD 47309} & \colhead{HD 95128} & \colhead{HD 71148}}
\startdata
$(B-V)$     & 0.619 & 0.651   & 0.616   & 0.623   & 0.606   & 0.625   \\
$(b-y)$     & 0.394 & 0.401   & 0.389   & 0.412   & 0.391   & 0.399   \\
$(Y-V)$     & 0.556 & 0.570   & 0.540   & \nodata & 0.550   & \nodata \\
$(V-S)$     & 0.546 & 0.540   & 0.550   & \nodata & 0.540   & \nodata \\
$(B_2-V_1)$ & 0.368 & 0.385   & 0.362   & \nodata & 0.363   & 0.367   \\
$(B_2-G)$   & 0.067 & 0.089   & 0.063   & \nodata & 0.057   & 0.070   \\
$t$         & 0.159 & 0.170   & 0.153   & \nodata & 0.149   & 0.159   \\
$(V-R_C)$   & 0.351 & 0.353   & \nodata & \nodata & \nodata & \nodata \\
$(V-I_C)$   & 0.682 & 0.688   & \nodata & \nodata & \nodata & \nodata \\
$(R_C-I_C)$ & 0.330 & 0.335   & \nodata & \nodata & \nodata & \nodata \\
$C(42-45)$  & 0.633 & 0.651   & 0.630   & \nodata & 0.629   & 0.634   \\
$C(42-48)$  & 1.650 & 1.671   & 1.640   & \nodata & 1.637   & 1.647   \\
$(B_T-V_T)$ & 0.689 & 0.736   & 0.711   & 0.686   & 0.733   & 0.692   \\
$(V-J_2)$   & 1.141 & \nodata & 1.188   & \nodata & 1.191   & 1.171   \\
$(V-H_2)$   & 1.396 & \nodata & 1.482   & \nodata & 1.433   & 1.451   \\
$(V-K_2)$   & 1.495 & \nodata & 1.552   & \nodata & 1.508   & 1.499   \\ 
$(V_T-K_2)$ & 1.562 & \nodata & 1.617   & \nodata & 1.578   & 1.563  \\     \hline
$\teff$\tablenotemark{1} & $5777\pm10$ & $5735\pm39$ & $5796\pm36$ & $5732\pm47$ & $5807\pm15$ & $5754\pm27$ \\ \hline
$\feh$\tablenotemark{2} & $+0.00$ & $+0.05$ & $-0.02$ & $+0.11$ & $+0.0$ & $-0.02$                   
\enddata
\tablenotetext{1}{The temperature of the Sun is the direct one, for the solar analogs the average of the temperatures from colors are given. Simple standard deviations are given as error bars.}
\tablenotetext{2}{$\feh=+0.0$ for the Sun, by definition. The metallicities of the solar analogs are from Soubiran \& Triaud (2004).}
\label{t:sun}
\end{deluxetable*}

The `closest ever solar twin', 18 Sco or HD~146233 (Porto de Mello \& da Silva 1997) appears to be $\sim40$~K cooler than the Sun. 
The other four stars in Table~\ref{t:sun} are those from the list of `Top Ten solar analogs in the ELODIE library' (Soubiran \& Triaud 2004) whose temperatures are around 5780~K. The remaining five stars of the `Top Ten' are in general cooler by about 100~K.

The range of $(B-V)$ colors for these five solar analogs is $0.61<(B-V)<0.65$, while the IRFM temperature scale suggests $(B-V)_\odot=0.62$, implying reasonable agreement considering photometric errors (which are around 0.01 mag) and metallicity effects. In general, the remaining colors of the solar analogs are also consistent with those we derived for the Sun with the IRFM $\teff$ scale, within photometric uncertainties.

As noted by Sekiguchi \& Fukugita (2000), the values of $(B-V)_\odot$ most often found in the literature range from 0.65 to 0.67, which are considerably larger than the present result. In their detailed study of the $(B-V)$ color-temperature relations they also find a bluer $(B-V)_\odot=0.626$. Therefore, in order to correctly place the Sun in the $\teff$ vs. $(B-V)$ plane, as defined by normal stars, $(B-V)_\odot$ should be bluer than what is usually quoted in the literature.

\subsection{Metallicity effects}

\begin{figure*}
\plotone{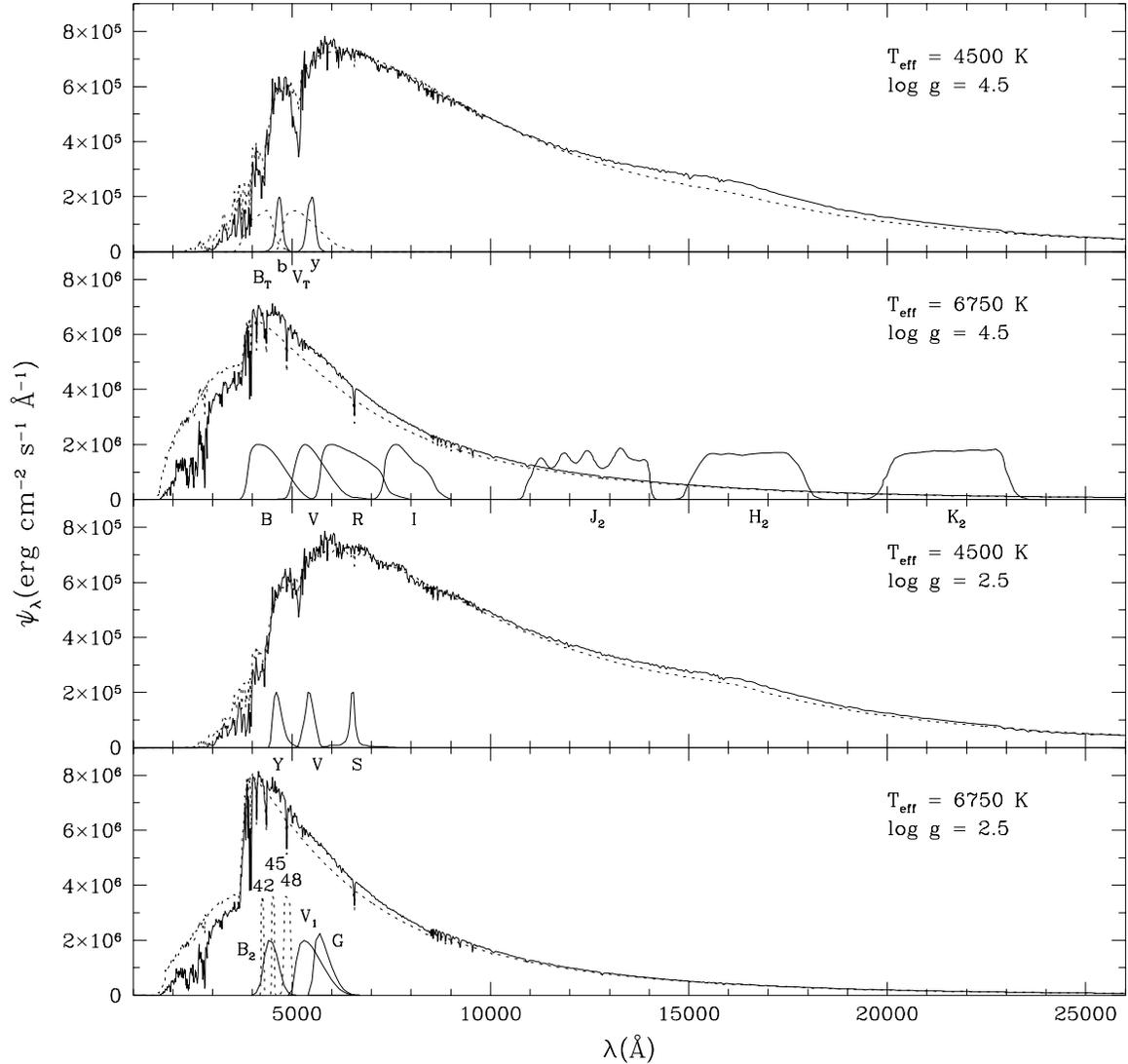} \caption{Spectral Energy Distributions (SEDs) from Kurucz models, as given by Lejeune et~al. (1997), for atmospheric parameters representative of cool ($\teff=4500$~K), hot ($\teff=6750$~K), giant ($\logg=2.5$), dwarf ($\logg=4.5$), metal-poor ($\feh=-2.5$, dotted lines), and metal-rich ($\feh=+0.0$, solid lines) stars. The filter transmission functions of interest to this work are also shown.} \label{fig:sedsF}
\end{figure*}

Metallicity has an important effect on the $\teff$ vs. color relations, particularly for the $(B-V)$ color (e.g. Cameron 1985, Mart\'{\i}nez-Roger et~al. 1992). Fig.~\ref{fig:sedsF} shows representative theoretical spectra of metal-poor and metal-rich dwarfs and giants at 4500~K and 6750~K. The transmission functions of the filters adopted in the present study are also shown. The figure is intended to be of use when trying to understand the effects of both $\feh$ and $\logg$ on the $\teff$ vs. color relations.

In general, at a given temperature, the colors get redder (larger) as more metals are present. The simplest explanation is that with more metals in the atmosphere, the UV and blue continuum is greatly reduced by line blanketing, with a corresponding increase of the red continuum due to flux redistribution, which results in the metal-rich stars being redder. This explanation is appropriate for colors constructed with filters measuring the fluxes in the ``blue'' (4000\AA$<\lambda<5500$\AA) and ``visual'' (4500\AA$<\lambda<6000$\AA) regions of the spectrum but must not be extended to other colors straightforwardly. The situation for the Cousins and Johnson-2MASS colors, for example, is substantially different.

As it can be clearly seen in Fig.~\ref{fig:dvk2} (dwarfs) and to a lesser extent in Fig.~\ref{fig:gvk2} (giants), the $(V-K_2)$ colors of metal-poor stars are indeed `bluer' (i.e. smaller) than the metal-rich stars at cool temperatures. However, for temperatures above 6000~K the metal-poor stars are \textit{redder}. This is consistent with the synthetic spectra shown in Fig.~\ref{fig:sedsF}. For $\teff=6750$~K and $\logg=4.5$, the flux in the $K_2$ band does not change substantially as $\feh$ is reduced from $+0.0$ to $-2.5$ dex. The flux in the visual region, on the other hand, is larger for the metal-rich spectrum due to the redistribution of the UV flux into the visual region. The net result is a larger $(V-K_2)$ for the metal-poor stars (assuming these theoretical spectra reasonably reproduce the real ones).

Given its relatively low dependence on $\feh$, $(V-K)$ is a very good temperature indicator. We caution, however, that according to our results this is valid only from 4800~K to 6000~K for main sequence stars. If $\feh$ is unknown, one may be tempted to use the calibrations for $(V-K)$ assuming $\feh=+0.0$ and 
adopt a solar metallicity $\teff$. If the temperature of the star results in 6000~K or more, however, and the star is metal-poor, for example $\feh\sim-3.0$, the temperature may be being underestimated by 200~K or even more. Thus, at these high temperatures, adopting a solar metallicity temperature is unacceptable for metal deficient dwarfs. Results from model atmospheres support this conclusion. The situation for cool dwarfs ($\teff<4500$~K) should not be considered conclusive, because although it is consistent with the models, the corresponding models themselves are not very reliable. For giants the effect of a wrongly adopted solar metallicity on $\teff$, for the very metal-poor stars, is $\sim$100~K or less. 


Ryan et~al. (1999) criticized Alonso et~al. (1996) $(b-y)$ calibration for dwarfs, arguing that it becomes unphysical below $\feh=-2.5$ for turn-off stars ($\teff\sim6750$~K), and attributed the effect to the quadratic $\feh$ term of the calibration formula (Eq. \ref{eq:theta}). Unphysical or not, it is the IRFM that produces higher temperatures for very metal-poor turn-off stars, an effect that is clearly seen not only in the $\teff$ vs. $(b-y)$ plane but also in the $\teff$ vs $(B-V)$ (Fig.~\ref{fig:dbv}), $(B_2-V_1)$, and $(B_2-G)$ planes. In fact, a large quadratic term (coefficient $a_5$ in Table~\ref{t:aD}) for these blue-optical colors may not be unphysical given the large blanketing effects in the blue-visual region (Fig.~\ref{fig:sedsF}). The effect is intrinsic to the IRFM and not a numerical artifact introduced by the calibrations.

\subsection{Surface gravity effects}

\begin{figure*}
\epsscale{0.8}
\plotone{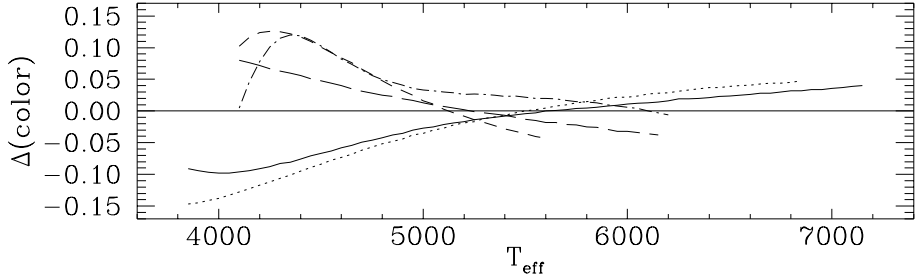} \caption{Color of a dwarf minus that of a giant at a fixed temperature as a function of $\teff$ for: $(B-V)$ (solid line), $(b-y)$ (dotted line), $(V-J_2)$ (dashed line), $(V-H_2)$ (long dashed line), and $(V-K_2)$ (dot-dashed line).} \label{fig:gravity}
\end{figure*}

The effects of surface gravity ($\logg$) on colors are illustrated in Fig.~\ref{fig:gravity}, where the difference between the color of a dwarf and a giant star of the same $\teff$ is plotted against $\teff$ for the $(B-V)$, $(b-y)$, $(V-J_2)$, $(V-H_2)$, and $(V-K_2)$ colors. A similar comparison for other colors can be found in Fig.~17 of RM04a.

For $\teff>6000$~K, decreasing $\logg$ has the effect of increasing the absolute value of the slope of the Paschen continuum (Fig.~\ref{fig:sedsF}, see Sect.~5.2 in RM04a for a physical explanation). Since $(B-V)$ and $(b-y)$ essentially measure this (negative) slope, their values for giants will be smaller (the slope becomes even more negative). This is consistent with the differences plotted in Fig.~\ref{fig:gravity} for $(B-V)$ and $(b-y)$.

Note that in the range 4800~K$<\teff<6000$~K the Johnson-2MASS colors are almost insensitive to $\logg$ (the effect is less than 0.05 mag), which makes them suitable for stars of unknown luminosity class, in addition to the fact that they are also nearly metallicity independent. The influence of gravity on colors at low temperatures is very complex and still difficult to understand (see Sect.~5.2 in RM04a).

\subsection{Giants in clusters}

In Part I, we derived the IRFM temperatures of a number of giants in the following clusters: M3, M67, M71, M92, 47~Tuc, NGC~1261, NGC~288, and NGC~362. When the calibrations for giants were performed, a preliminary field calibration was first constructed, and then only the clusters for which the internal cluster $\teff$ scale was consistent with the field $\teff$ scale were included. This procedure not only reduced the risk of systematic errors introduced by errors in the metallicity and reddening corrections of the clusters, but mainly avoided errors in the stellar photometry. It is not surprising, then, that the $\teff$ scale for the DDO colors includes all the clusters (see Fig.~18 in RM04a). The photometry is highly accurate, and the metallicity effects on DDO colors strong. This result confirms that the $\teff$ scale of giants in the field and that of giants in clusters is the same, and that the metallicity and reddening scales of the clusters were well determined.

However, there are still small discrepancies in the metallicities of globular clusters, e.g. for M71 Ram\'{\i}rez et~al. (2001) give $\feh=-0.77$, Sneden et~al. (1994) $\feh=-0.79$, and Carreta \& Gratton (1997, hereafter CG97) $\feh=-0.70$; for NCG~362 Shetrone \& Keane (2000) suggest $\feh=-1.33$ and CG97 $\feh=-1.15$; while for M92 Sneden et~al. (2000) obtain $\feh=-2.37$, King et~al. (1998) $\feh=-2.52$, and CG97 $\feh=-2.16$. The values we adopted are mainly those given by Kraft \& Ivans (2003), which are in reasonable agreement with the mean of the different values cited in the literature.

\begin{figure}
\epsscale{1.}
\plotone{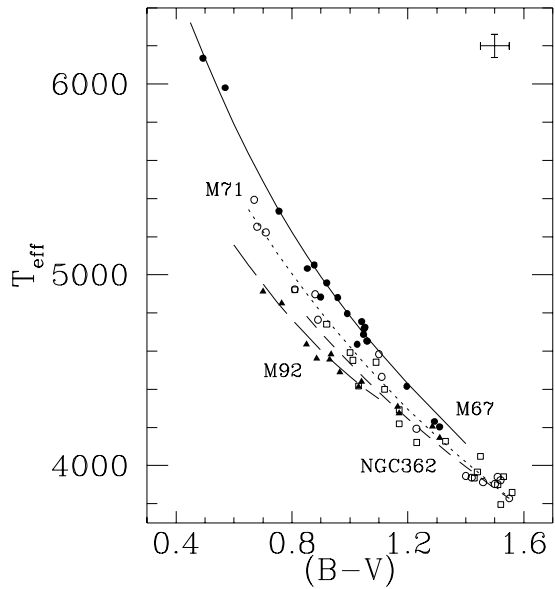} \caption{$\teff$ vs. $(B-V)$ relations for giants in the open cluster M67 (filled circles) and in the globular clusters M71 (open circles), NGC~362 (squares), and M92 (triangles). Solid, dotted, dashed, and long-dashed lines correspond to our calibrations for $\feh=-0.08$, $\feh=-0.80$, $\feh=-1.30$, and $\feh=-2.40$, respectively, which are approximately the mean metallicities adopted for the clusters. Typical errors bars (0.05 mag and 60~K) are shown to the upper right corner of the figure.} \label{fig:escalag}
\end{figure}

The $\teff$ vs. $(B-V)$ calibration does not include M3, NGC~288, NGC~1261, and 47~Tuc due to large photometric uncertainties. The photometry for the other clusters reproduce the $(B-V)$ $\teff$ scale very well (Fig.~\ref{fig:escalag}). Even for the RGB tip of M3 we found good agreement when the best available photometry of three bright stars was used, but the very large photometric errors of the fainter giants produced a large disagreement between the two scales.

\begin{figure}
\epsscale{1.}
\plotone{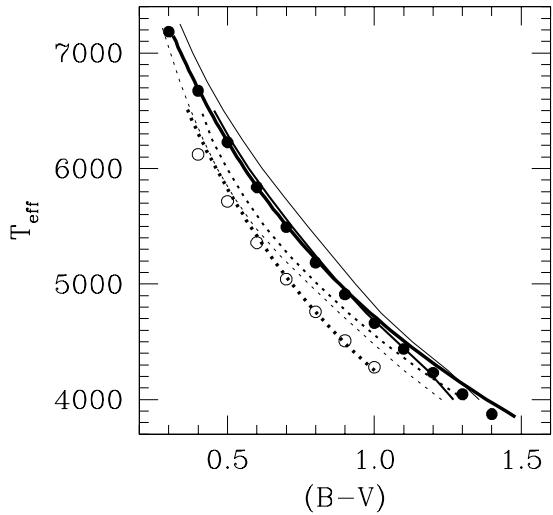} \caption{$\teff$ vs. $(B-V)$ relation for main sequence stars of $\feh=+0.0$ (solid lines and filled circles) and $\feh=-2.5$ (dotted lines and open circles) according to: this work (thick black lines), Alonso et~al. (circles), Bessell (thin cyan lines), and Houdashelt et~al. (magenta lines).} \label{fig:comp1}
\end{figure}

\section{Comparison with other studies} \label{sect:comparison}

A considerable body of work exists on the $\teff$ scale of F, G, and K stars; some of these works have been discussed in Part~I (Sect.~5.3). The IRFM $\teff$ scale has been compared with the results of several groups and has been found, in general, in reasonable agreement with them (AAM, MR03, RM04a). The reader is referred to AAM, MR03, and RM04a to see how the present IRFM $\teff$ scale specifically compares to other results. Here we will proceed to show that the present work is an update of AAM work and our earlier extensions. We also revisit the comparisons with the synthetic colors derived from Kurucz models (M.~S.~Bessell 2004, private communication) and MARCS models (Houdashelt et~al. 2000).

\begin{figure}
\epsscale{1.}
\plotone{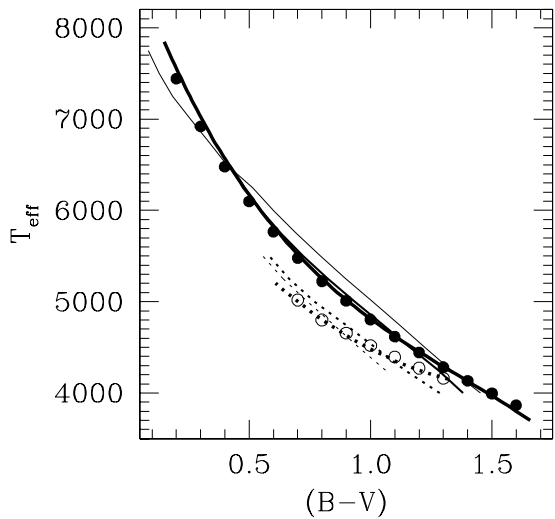} \caption{$\teff$ vs. $(B-V)$ relation for giant stars of $\feh=+0.0$ (solid lines and filled circles) and $\feh=-2.0$ (dotted lines and open circles) according to: this work (thick black lines), Alonso et~al. (circles), Bessell (thin cyan lines), and Houdashelt et~al. (magenta lines).} \label{fig:comp2}
\end{figure}

In Figs.~\ref{fig:comp1} and \ref{fig:comp2} we compare our $\teff$ vs. $(B-V)$ relation with that given by AAM for dwarfs of $\feh=+0.0$ and $\feh=-2.5$, and giants of $\feh=+0.0$ and $\feh=-2.0$, respectively. These particular values of $\feh$ were adopted to maximize the ranges in common. The $\feh$ difference is also large enough as to easily distinguish the solar-metallicity $\teff$ vs. color relation from the metal-poor one. The theoretical calibrations of Bessell and collaborators (see e.g. Bessell et~al. 1998), as well as that of Houdashelt et~al. (2000), are also shown (for the dwarf comparison, colors for $\logg=4.5$ were adopted, for giants we used the colors for $\logg=2.5$).

\begin{figure*}
\epsscale{0.8}
\plotone{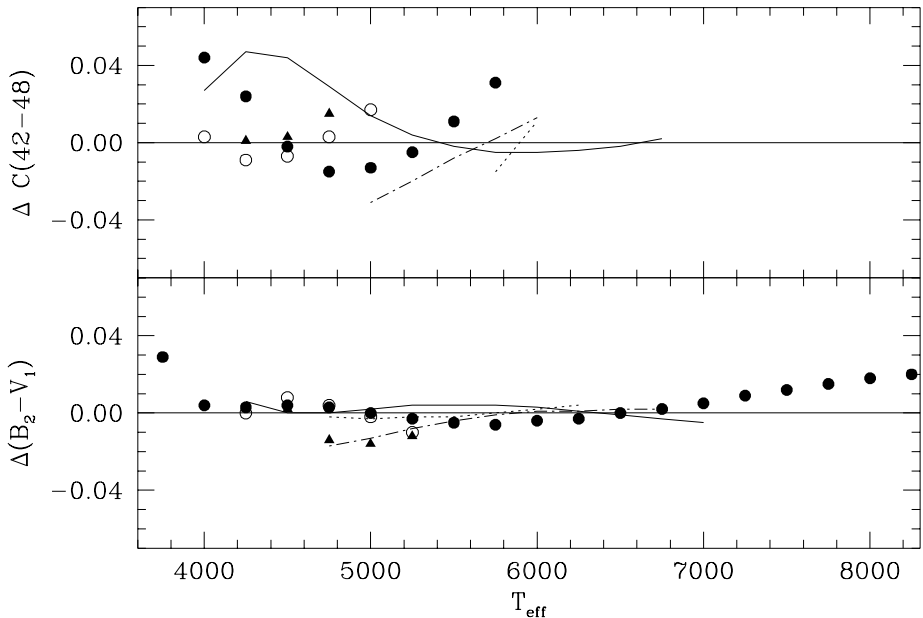} \caption{Color difference at a given temperature between the $\teff$ scale derived in this work and that derived in MR03 and RM04a, which is based on Alonso et~al. temperatures. Solid, dash-dotted, and dotted lines correspond to $\feh=+0.0$, $\feh=-1.0$, and $\feh=-2.0$ main sequence stars, respectively. Filled circles, open circles, and triangles correspond to $\feh=+0.0$, $\feh=-1.0$, and $\feh=-2.0$ giants, respectively.} \label{fig:comp3}
\end{figure*}

\begin{figure*}
\epsscale{0.9}
\plotone{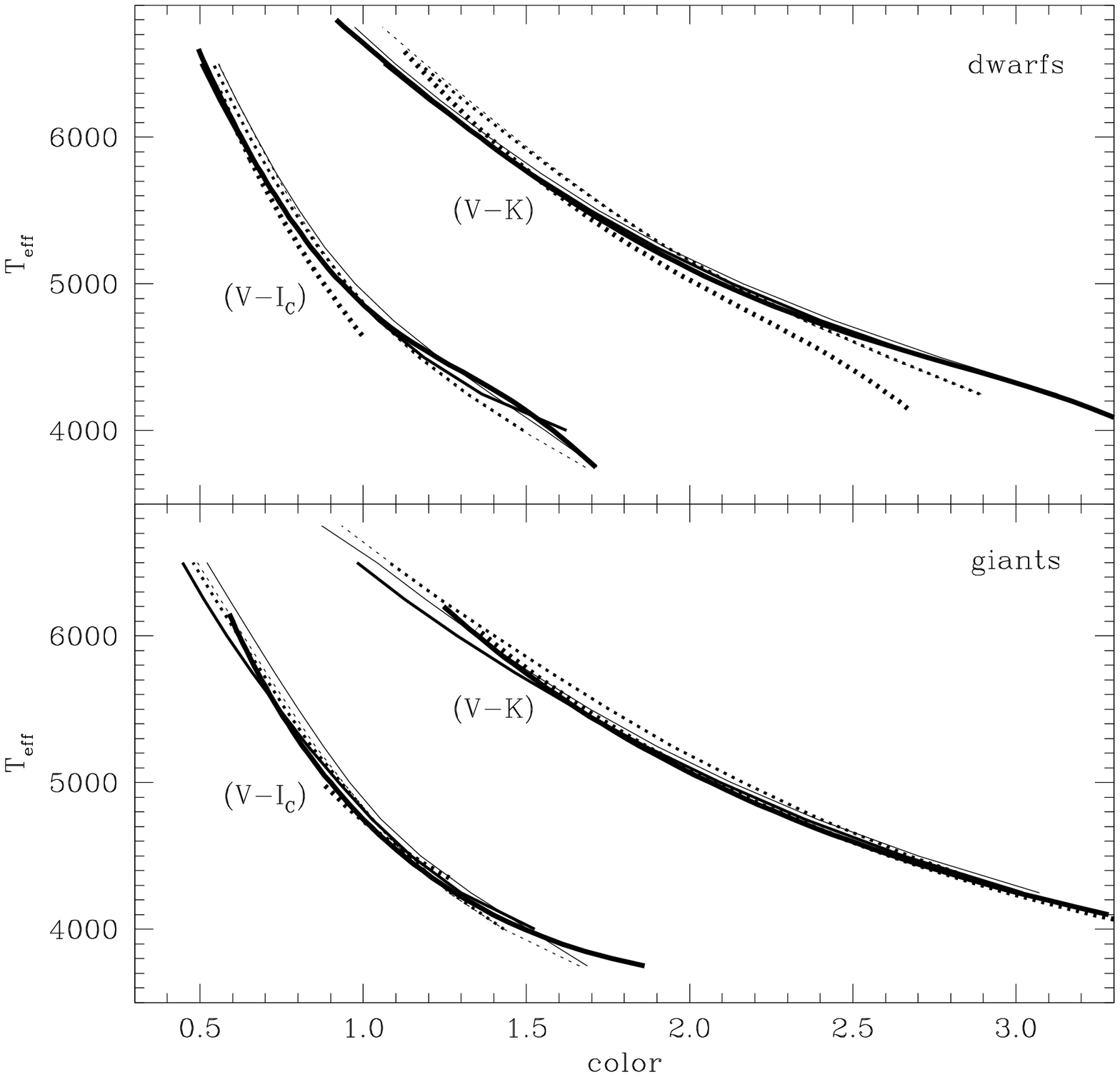} \caption{$\teff$ vs. $(V-I_C)$ and $(V-K)$ relations for dwarf (upper pannel) and giant (lower pannel) stars of $\feh=+0.0$ (solid lines) and $\feh=-2.0$ (dotted lines) according to: this work (thick black lines), Bessell (thin cyan lines), and Houdashelt et~al. (magenta lines).} \label{fig:comp4}
\end{figure*}

Our $\teff$ vs. $(B-V)$ relations are essentially the same as those of AAM. There are, however, small departures at low temperatures ($<4500$~K for dwarfs, $<4000$~K for giants) for solar metallicity, and around $\teff=6000$~K for the metal-poor end. In the latter case, for a fixed temperature, the colors from our $\teff$ scale are redder by about 0.03 mag. Note that AAM relations for dwarfs are almost parallel, i.e. the effect of $\feh$ appears to be color (or temperature) independent. The synthetic colors obtained from both Kurucz and MARCS models, however, show a quite complicated behaviour, the relations for solar metallicity and $\feh=-2.5$ are closer both at the cool and hot extremes. The trends predicted by model atmospheres at $\teff>6000$~K are consistent with our results. On the other hand, the synthetic colors seem to be failing to reproduce the strong metallicity effects for cool ($\teff<5000$~K) dwarfs. According to our IRFM $\teff$ scale, at $(B-V)=1.0$ the temperature for $\feh=0.0$ is about 400~K higher than that at $\feh-2.5$, but Houdashelt et~al. suggest a $\Delta\teff$ of only about +100~K.

AAM temperatures for giants are slightly cooler than ours for $\teff\gtrsim6500$~K but they are in an almost perfect agreement with ours everywhere else. According to the IRFM, the relations of $\feh=+0.0$ and $\feh=-2.0$, in the giant $\teff$ vs. $(B-V)$ plane, get closer as $\teff$ is reduced. This behaviour is not in agreement with the theoretical calculations, which suggest an almost constant $\feh$ effect (Fig.~\ref{fig:comp2}). The solar metallicity prediction of Houdashelt et~al. is in excellent agreement with the present results except at the cool end.

In Fig.~\ref{fig:comp3} we compare the $C(42-48)$ and $(B_2-V_1)$ colors as a function of $\teff$ and $\feh$ for both dwarf and giant stars, as obtained in this work, with those from our earlier work (MR03 and RM04a). The main difference is the updated temperatures of the present work, of course; MR03 and RM04a are based on AAM temperatures. Also, no polynomial corrections were performed in MR03. The comparison shows that the two $\teff$ scales differ by less than 0.05 mag in $C(42-48)$ and 0.03 mag in $(B_2-V_1)$. These differences are well within the photometric errors, systematic errors in the previous calibrations, and errors in the temperatures. The differences do not exhibit any significant trend with either $\teff$ or $\feh$. The differences are larger for the DDO color calibration, which is due to the lower number of stars defining the $\teff$ vs. $C(42-48)$ relation as compared to the $\teff$ vs. $(B_2-V_1)$ relation. The calibration for $(B_2-V_1)$ is obviously more robust, but even in the case of $C(42-48)$ the differences are not large.


A comparison of our $\teff$ vs. $(V-I_C)$ and $(V-K)$ relations with those given by Bessell and Houdashelt et~al. is illustrated in Fig.~\ref{fig:comp4}. Predictions of solar metallicity and $\feh=-2.0$ are shown. The $(V-K)$ colors of Bessell are in the Bessell \& Brett (1988) system, those by Houdashelt et~al. are in the Johnson-Glass system; and ours are in the Johnson-2MASS system. Their exact values may be somewhat different, depending on the color, but the metallicity effects on the colors should be essentially the same. In any case, the Bessell \& Brett $(V-K)$ colors, for example, only need to be shifted by 0.04 mag to be transformed into the Johnson-2MASS system (Carpenter 2001). The comparison for the $(V-K)$ colors is thus still meaningful.

At solar metallicity the $\Delta\teff/\Delta\mathrm{color}$ gradients are in reasonable agreement with our dwarf calibrations above 4500~K and with the giant calibrations below 5500~K. The effect of the metallicity is remarkably different between the two theoretical results for the $(V-I_C)$ color at high temperatures: while Houdashelt et~al. colors tend to be redder with a lower $\feh$, Bessell suggests a bluer color. The latter is in better agreement with our results, but only differentially, as a shift of about 150~K may be required to match the two $\teff$ scales. Regarding $(V-K)$, the general trends are very similar at low temperatures, although the IRFM suggests a very strong metallicity effect that makes the cool dwarfs very blue. They are bluer also according to the theoretical results but the effect there is not very strong. For $\teff>6000$~K both sets of synthetic colors agree in that the metal-poor stars are redder, a result discussed in \S4.4. Note, however, that the synthetic $(V-K)$ colors are even more metallicity dependent.

Kurucz and MARCS colors are thus in reasonable qualitative agreement with the IRFM, but the synthetic
colors may be failing to predict the detailed metallicity dependence. Even the solar metallicity colors are not very well reproduced by model atmospheres and synthetic colors, since the predicted colors require not only a zero point correction, but a correction that depends also on spectral type (see e.g. Sect. 3.2.2 in Houdashelt et al.~2000). In particular, the calibration of Houdashelt et~al., while in excellent agreement with the IRFM for solar metallicity stars, shows differences as large as $\pm200$~K for stars with $\feh=-2$, but note that it depends on the color being compared, metallicity, evolutionary stage and spectral type. For example, for an F metal-poor ($\feh=-2$) dwarf, the agreement between the IRFM and Houdashelt et al. $(V-K)$ colors is very good, but for cooler stars the temperatures derived from Houdashelt et al. are systematically higher, and at $(V-K)=2.5$, Houdashelt et al. temperatures are higher than those from the IRFM by 200~K. These differences become especially important for samples covering a large range in $\teff$; for example, when studying small abundance variations from the turn-off to the RGB tip of globular clusters, spurious variations could be found for stars of different evolutionary stages. Fortunately, the problem is alleviated when abundances with respect to iron are reported ([X/\ion{Fe}{1}]), unless the lines of the element X have a temperature dependence very different from those of \ion{Fe}{1}.

\section{Conclusions} \label{sect:conclusions}

Based on a large number of main sequence and giant stars with temperatures determined with the infrared flux method (IRFM), a set of homogeneously calibrated temperature vs. color relations is provided. The calibrations include the effect of $\feh$ and residual corrections to avoid systematic effects introduced by the calibration formulae, with the aim of reproducing the effects of spectral features that cannot be taken into account by the initial fits.

The calibrations have been tested with a sample of stars with known direct and IRFM temperatures. Usually, more than eight colors were adopted to derive a photometric temperature. The comparison of photometric temperatures with those from the IRFM shows excellent agreement, with a dispersion fully consistent with the errors in the IRFM temperatures. Thus, we have shown that the calibrations do not introduce any systematic error. When compared to the direct temperatures, not only is good agreement found, but we have also shown that the zero point of the IRFM temperature scale is in agreement with the absolute zero point, to a level of about 10~K.

The colors of the Sun, as determined from the calibrations for a star of $\teff=5777$~K and $\feh=+0.0$, are presented and compared with those of five solar analogs. A very good agreement is found considering the photometric uncertainties, which makes these colors useful to the search of solar twins in various surveys, particularly Hipparcos-Tycho and 2MASS.

Metallicity and surface gravity effects on the IRFM $\teff$ scale may be reasonably understood with the help of theoretical spectra, although the situation for K stars is still difficult to address. For stars with $\teff\gtrsim6000$~K, the adoption of a solar metallicity temperature derived from the $(V-K)$ color for a metal-poor star is unacceptable, as systematic errors of the order of 200~K may be introduced.

Provided that the photometry is accurate, good agreement is found when comparing the temperature scales of giants in the field and giants in clusters. Thus, the IRFM $\teff$ scale is the same for both field and cluster giants.

As expected, our $\teff$ scale closely resembles the one that is based on the Alonso et~al. temperatures (our work is largely based on their study). The present results, however, have a better coverage of the atmospheric parameter space and are more reliable: better and updated input data was adopted, and the calibrations were carefully performed to avoid the introduction of numerical sources of error. A number of problems of interest to stellar evolution and the chemical evolution of the Galaxy depend on the assumptions made in color vs. $\teff$ relations. Our calibrations will permit these problems to be tackled with greater confidence.

\acknowledgments{JM thanks partial support from NSF grant AST-0205951 to JG Cohen. We thank M.~S.~Bessell for providing colors for a more complete set of Kurucz models, I.~Ivans for improving the English language and style of the manuscript, and the anonymous referee for comments and suggestions that helped to improve the paper. This publication makes use of data products from 2MASS, which is a joint project of the University of Massachusetts and IPAC/Caltech, funded by NASA and the National Science Foundation; and the SIMBAD database, operated at CDS, Strasbourg, France.}

\end{document}